%% file: main.tex
\documentclass[preprint,12pt]{elsarticle}
\usepackage[left=2.5cm,top=3cm,right=2.5cm,bottom=3cm,bindingoffset=0.5cm]{geometry}

\usepackage{lineno,hyperref}
\usepackage{amsmath} 
\usepackage{graphics} 
\usepackage{subfig} 
\usepackage{amssymb}

\def\xseff{\sigma_{\textit{eff}}}

\journal{Nuclear Instruments and Methods in Physics Research Section A}

\begin{document}

\begin{frontmatter}

\title{Measurement of the neutron flux at spallation sources using multi-foil activation}

\author[mib,infnmib]{Davide Chiesa\corref{mycorrespondingauthor}}
\cortext[mycorrespondingauthor]{Corresponding author}
\ead{davide.chiesa@mib.infn.it}

\author[mib,infnmib]{Massimiliano Nastasi}
\author[ral]{Carlo Cazzaniga}
\author[ifp,infnmib,mib]{Marica Rebai}
\author[fermi,TorVergata,UniLondon]{Laura Arcidiacono}
\author[infnmib]{Ezio Previtali}
\author[mib,ifp,infnmib]{Giuseppe Gorini}
\author[ral]{Christopher D. Frost}

\address[mib]{Physics Department ``G. Occhialini" of Milano-Bicocca University, piazza della Scienza 3, 20126 Italy}
\address[infnmib]{INFN section of Milano-Bicocca, piazza della Scienza 3, 20126 Italy}
\address[ral]{ISIS Facility, STFC, Rutherford Appleton Laboratory, Didcot OX11 0QX, UK}
\address[ifp]{IFP-CNR, Via Cozzi 53, 20125, Milano, Italy}
\address[fermi]{Museo Storico della Fisica e Centro Studi e Ricerche "Enrico Fermi", Piazza del Viminale 1,00184 Roma, Italy}
\address[TorVergata]{Universit\'a degli Studi di Roma ``Tor Vergata", NAST Center, Via della Ricerca Scientifica 1, Rome, 00133 Italy}
\address[UniLondon]{University College of London - Institute of Archaeology, 31-34 Gordon Square London WC1H 0PY, United Kingdom} 

\begin{abstract}
Activation analysis is used in this work to measure the flux of a fast neutron beamline at a spallation source over a wide energy spectrum, extending from thermal to hundreds of MeV. \\
The experimental method is based on the irradiation of multiple elements and measurements of activation $\gamma$-lines using a High Purity Germanium detector. \\
The method for data analysis is then described in detail, with particular attention to the evaluation of uncertainties. The reactions have been chosen so to cover the whole energy range, using mainly (n,$\gamma$) for thermal and epithermal neutrons, and threshold reactions for the fast region.  The variety of these reactions allowed for the unfolding of the neutron spectrum, using an algorithm based on a Bayesian statistical model, and limited correlations have been found between the energy groups.

\end{abstract}


\end{frontmatter}


\section{Introduction}
\input{introduction}

\section{Methodology} \label{sec:method}
\input{methodology}

\section{Experiment description} \label{sec:experiment}
\input{experiment}

\section{Data analysis and results} \label{sec:results}
\input{results}

\section{Conclusion}
\input{conclusion}

\section*{References}

\bibliography{Bibliography}

\end{document}

%% file: introduction.tex
Spallation sources produce neutrons via the interaction of accelerated high energy protons, typically in the range of hundreds of MeV to few GeV, with heavy target nuclei~\cite{russell1995introduction}. The result is a white neutron spectrum that extends over many orders of magnitude up to the energy of the accelerator~\cite{weisskopf1937statistics}. \\
The shape of the spectrum on experimental beamlines can be very different according to the specific design of the source and target station. Some major spallation sources are dedicated mainly to condense matter physics with cold, thermal or epithermal neurons, and therefore use moderators and reflectors~\cite{wilson1995guided,mason2006spallation,klinkby2016neutron}. Other sources are dedicated to fast neutrons applications and therefore use the direct products of the spallation target~\cite{lisowski1990alamos,guerrero2013performance,prokofiev2009anita}.  In this paper we propose the application of a method based on activation foil measurements and Bayesian unfolding to determine the neutron flux spectrum over this wide energy range. \\
The measurements presented in this paper have been performed on ChipIr~\cite{frost2009new}, a new beamline of the ISIS spallation source at the Rutherford Appleton Laboratory, UK. The ISIS synchrotron accelerates 700~MeV or 800~MeV protons that are collided onto a tungsten target. The measurements presented in this paper have been performed with the accelerator running at 700~MeV. ChipIr is a fast neutron beamline, specifically dedicated to the irradiation of microelectronics. The case stimulates particular interest because it is the first time that a fast neutron beamline has been built on a target station specifically designed for thermal neutrons. The case of ChipIr is to be compared with other  facilities with fast neutron targets dedicated to the irradiation of microelectronics, TRIUMF~\cite{blackmore2009development,blackmore2014intensity}, LANSCE~\cite{wender2016neutron}, TSL~\cite{prokofiev2013cup,prokofiev2009anita}, and with the VESUVIO beamline at ISIS~\cite{andreani2008facility,bedogni2009characterization,cazzaniga2015telescope}.
The innovation of the proposed method, with respect to what was done on other beamlines, lays mainly in the large variety of measured reactions, which are used in the unfolding analysis that, as it will be shown, gives a result with little dependence on the guess spectrum shape.

The method of measurement is extensively discussed in the next section. It is interesting to discuss here the choice of activation foils, a passive method, with respect to other techniques based on active detectors. This is motivated mainly by the following two reasons. (1) The ISIS source is pulsed at 10~Hz on the second target station (the target used by ChipIr) and the pulse has a double structure with two bunches 70~ns wide and 360~ns apart. This complex time structure does not allow for spectroscopy with time of flight techniques for high energy neutron ($E_n > 10$~MeV)~\cite{wender1993fission}. Furthermore, high instantaneous fluxes can result in a strong pile-up for many active detectors. Of course a passive method is not affected by these complications. (2) Flux and spectroscopy measurements of a neutron field are dependent on detector response functions and cross sections that are not always well known above 20 MeV. This is true for both active and passive methods, but the activation foil technique relies directly to cross sections verified at best by the scientific community rather than to more complex Monte Carlo simulations of response functions. \\
However, active detectors will also play an important role for ChipIr. They are needed for the following purposes. (1) Real time beam monitoring and dosimetry are performed with solid state detectors (diamond and silicon) and fission chambers~\cite{cazzaniga2016characterization,rebai2016time}.  (2) Beam mapping and profiling is performed with diamond~\cite{cazzaniga2017charge} and GEM detectors~\cite{croci2015gem}. (3) Spectroscopy with enhanced energy resolution is being investigated using the telescope proton recoil technique~\cite{cazzaniga2014thin}.

In this paper the unfolding methodology, based on multi-foil activation data, is presented in Section~\ref{sec:method}. This will be followed, in Section~\ref{sec:experiment}, by a description of the experimental method used at ChipIr, including the irradiation procedure and gamma spectroscopy. Finally, data analysis and results are presented in Section~\ref{sec:results}, including a discussion of the uncertainties that propagate from measurements, cross section data, and a systematic due to the guess spectrum.

The results of the presented measurements are of interest for the characterization of new fast neutron beamlines (like ChipIr), but also, more in general, for other beamlines at spallation sources (thermal or epithermal), and other kind of neutron sources like nuclear reactors.

%% file: methodology.tex
The neutron flux measurement with the neutron activation technique consists of irradiating samples with a known amount of target nuclei ($N$) and then measuring the \textit{activation rate} $R$, i.e. the number of reactions per unit time that produce a radioisotope in the sample. The relation between the neutron flux $\varphi$ and the activation rate is:
\begin{equation}
R=N\int\varphi(E)\sigma(E)dE
\label{eq:R}
\end{equation}
where $\sigma(E)$ is the activation cross section as function of neutron energy.

If the energy spectrum of the neutron flux is well known, the \textit{effective cross section} $\xseff$ is calculated as follows:
\[\xseff = \frac{\int\varphi(E)\sigma(E)dE}{\int\varphi(E)dE}  \] 
and, thus, the \textit{neutron flux intensity} $\Phi = \int\varphi(E)dE$ can be calculated from the measurement of only one reaction~\cite{AbsoluteFlux, FluxDistribution}:
\[\Phi = \frac{R}{N \xseff}\]

When the energy spectrum is not known, the neutron flux can be measured combining the activation data of different reactions. Depending on the cross section, each reaction is induced in different proportions by neutrons belonging to different parts of the energy spectrum. The radiative capture $(n,\gamma)$ reactions, characterized by cross sections with resonances at different neutron energies, can be used to measure the neutrons in the thermal and epithermal range. The threshold reactions allow to measure the fast neutrons with energies $\gtrsim$1~MeV.
Therefore, by properly choosing the set of activation reactions, a wide energy range of the neutron flux can be measured. 
As shown in~\cite{BayesianSpectrum}, the neutron flux spectrum can be evaluated with the following unfolding method.

A proper binning is chosen to subdivide the energy spectrum into $n$ \textit{flux groups}:
\[\phi_{i}\equiv\intop_{E_{i}}^{E_{i+1}}\varphi(E)dE\] 
For each activation reaction $j$, Eq.~\ref{eq:R} is rewritten as function of the flux groups $\phi_{i}$:
\begin{equation}
R_{j}=N_{j}{\sum_{i=1}^{n}}\sigma_{ij}\phi_{i}
\label{eq:system}
\end{equation}
where $\sigma_{ij}$ is the effective cross section in the energy bin of $i$-th group.\\

Eq.~\ref{eq:system} is a system of linear equations whose unknown variables are $\phi_{i}$. 
This system is solved with a statistical approach that allows to keep into account the experimental uncertainties affecting the parameters $R_j$ and $\sigma_{ij}$. 
In order to select only the physical solutions of the problem, a Bayesian statistical model is defined using positive definite \textit{probability distribution functions} (PDF) as \textit{priors} for $\phi_i$. 
The solutions for $\phi_{i}$ variables are determined by sampling the \textit{joint posterior} PDF $p(\phi_{i}|R_j,\sigma_{ij})$, i.e. the probability of all flux group intensities given the experimental data (and uncertainties) of activation rates and effective cross sections.
According to Bayes' theorem, the \textit{posterior} PDF is proportional to the product of \textit{likelihoods} and \textit{priors} distributions. 
The experimental data and uncertainties of $R_j$ and $\sigma_{ij}$ are used to define Gaussian likelihoods, while positive definite uniform distributions can be used as priors for $\phi_i$. 
The JAGS tool~\cite{JAGS} for Bayesian analysis allows to define a statistical model of the problem and to sample the \textit{posterior} PDF using Markov Chains Monte Carlo (MCMC) simulations~\cite{Gelman}. 
Once the \textit{posterior} is sampled, $\phi_{i}$ mean values and uncertainties are obtained from the marginalized distributions of the joint PDF. Moreover, the correlations between the $\phi_{i}$ variables are directly computed from the sampling of the joint PDF.

For as it is conceived, this technique requires a \textit{guess} spectrum to calculate the $\sigma_{ij}$ effective cross sections in each group of neutron energies.
Therefore, the $\sigma_{ij}$ values and, thus, the results of $\phi_{i}$ have some degree of dependence on the \textit{intra-group} spectrum shape used for $\sigma_{ij}$ calculation.
However, we stress that the guess spectrum choice does not put any constraint on the intensities of $\phi_{i}$.
Obviously, the more the groups are, the smaller will be the dependence of the results on the guess spectrum.
However, the number of groups that can be used for a successful unfolding of the neutron flux spectrum is strictly related to the properties of the cross sections of the activation reactions.

\begin{figure}[!htb]
\centering
\subfloat{\includegraphics[width=0.49\textwidth]{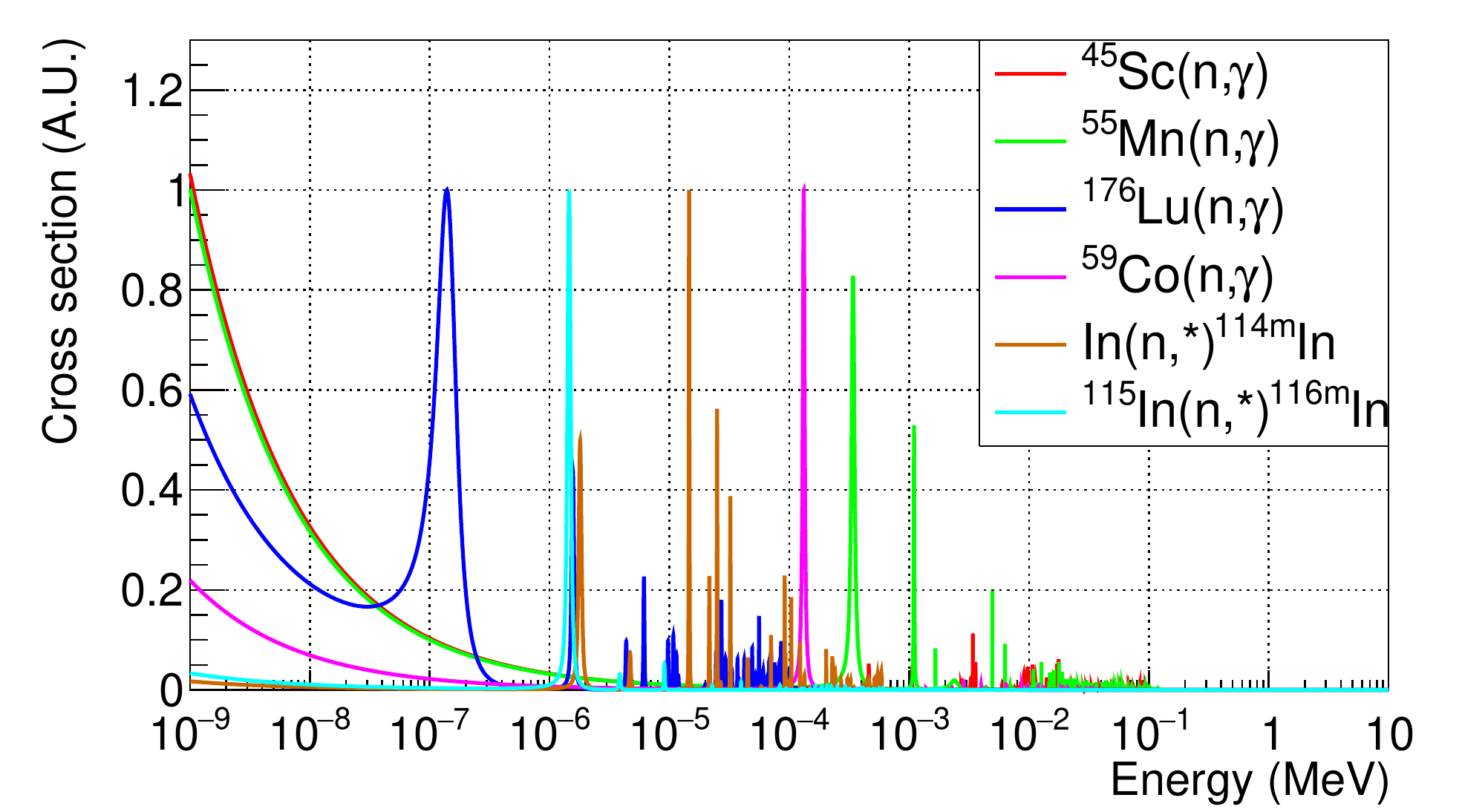}}
\subfloat{\includegraphics[width=0.49\textwidth]{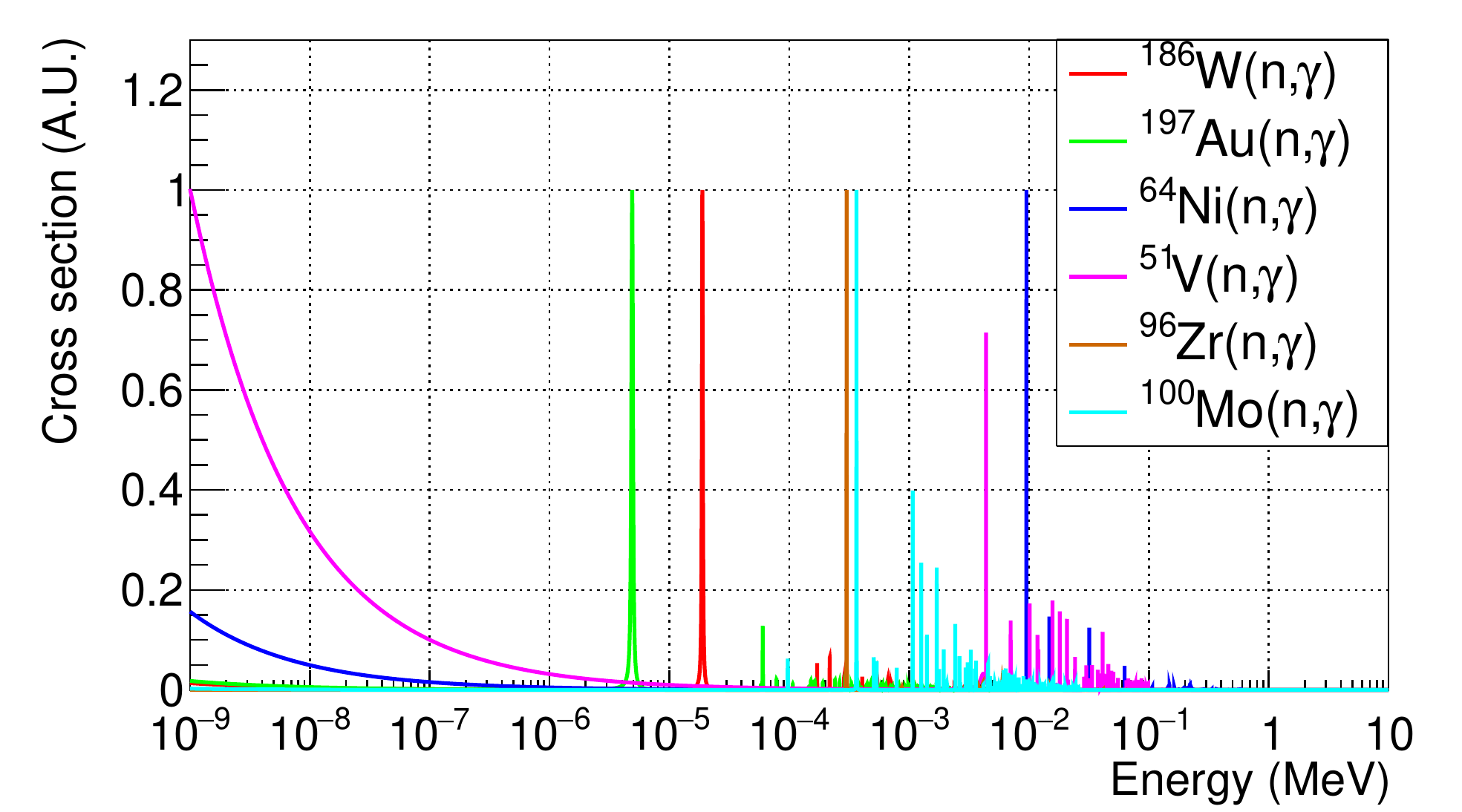}}
\\
\subfloat{\includegraphics[width=0.49\textwidth]{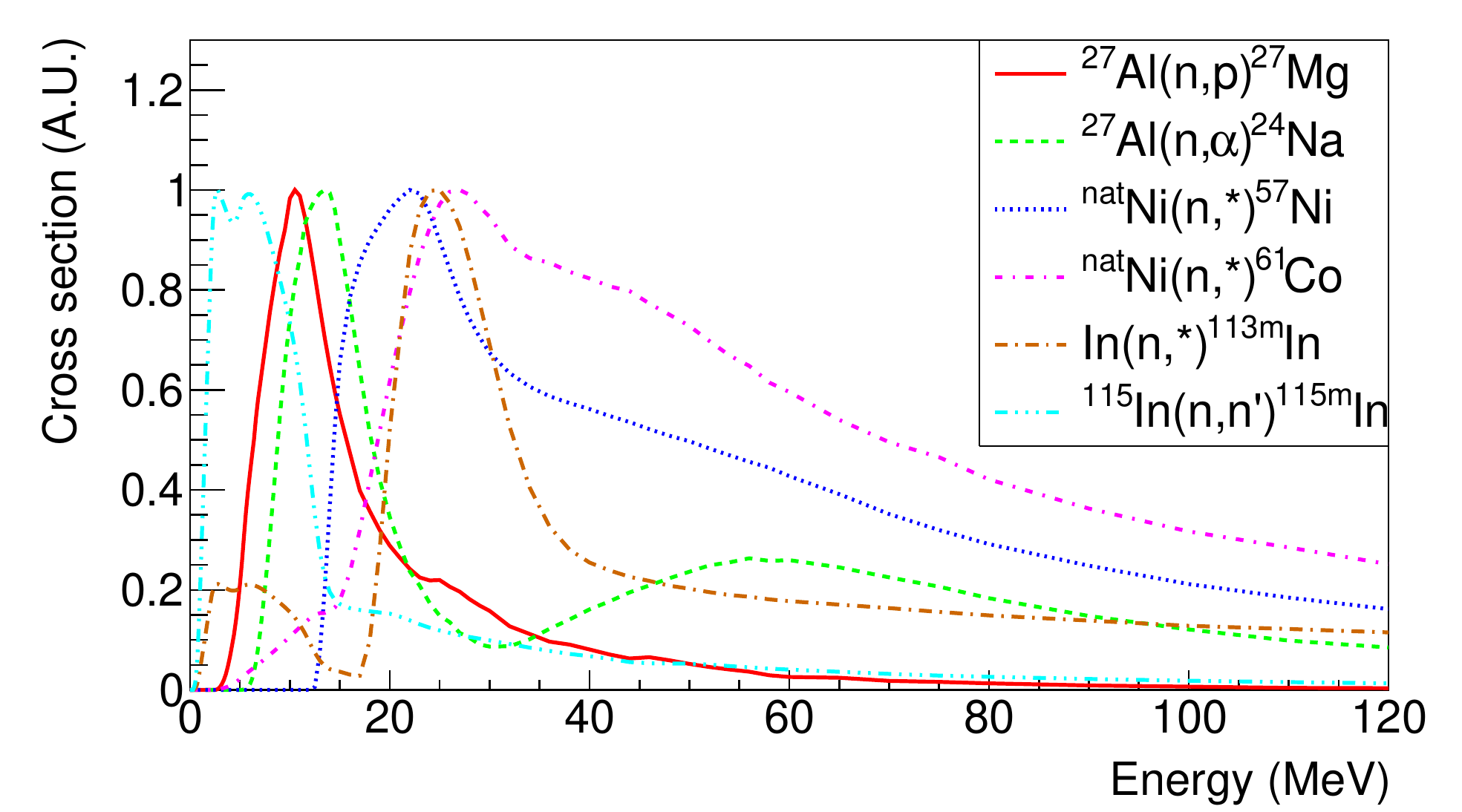}}
\subfloat{\includegraphics[width=0.49\textwidth]{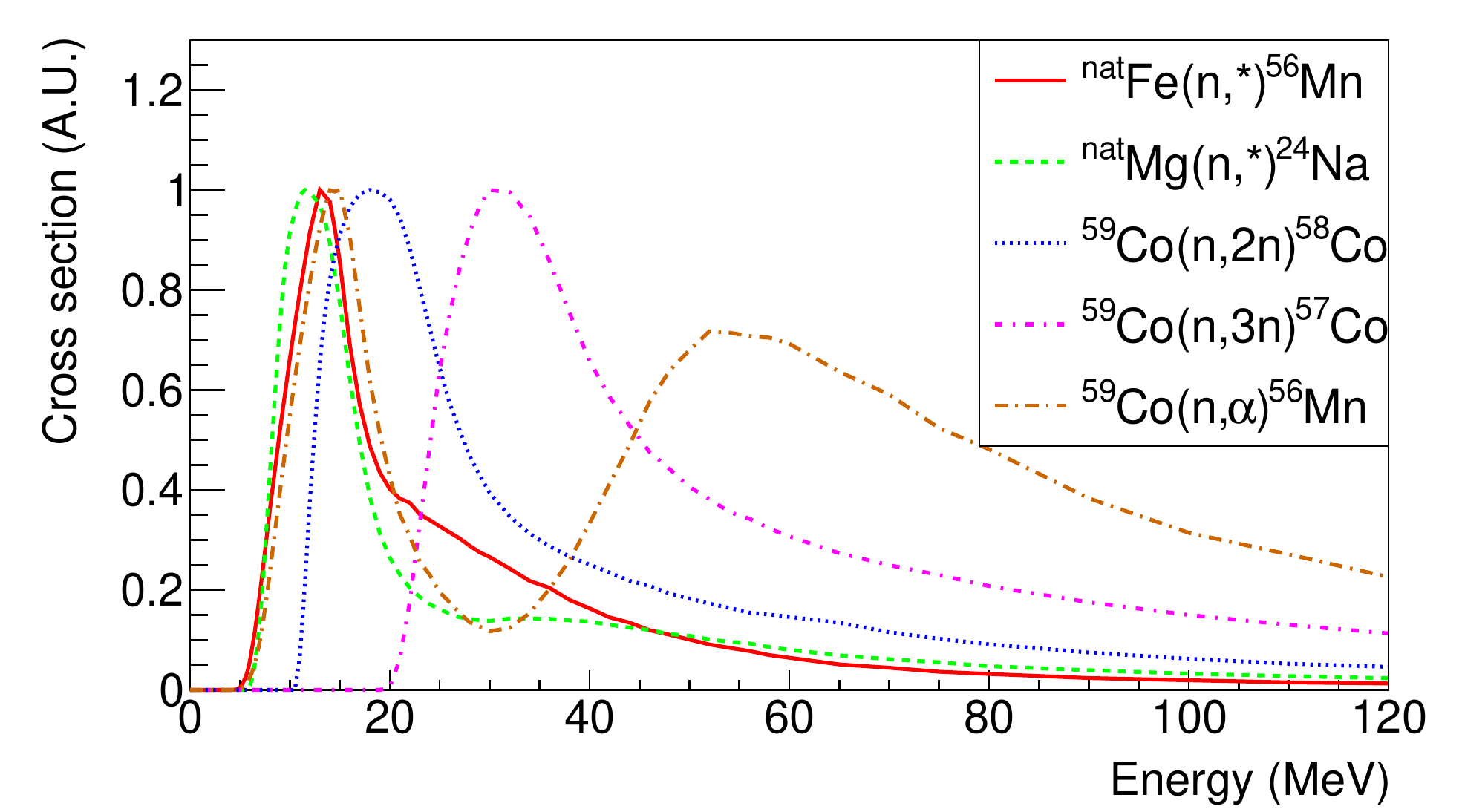}}
\\
\subfloat{\includegraphics[width=0.49\textwidth]{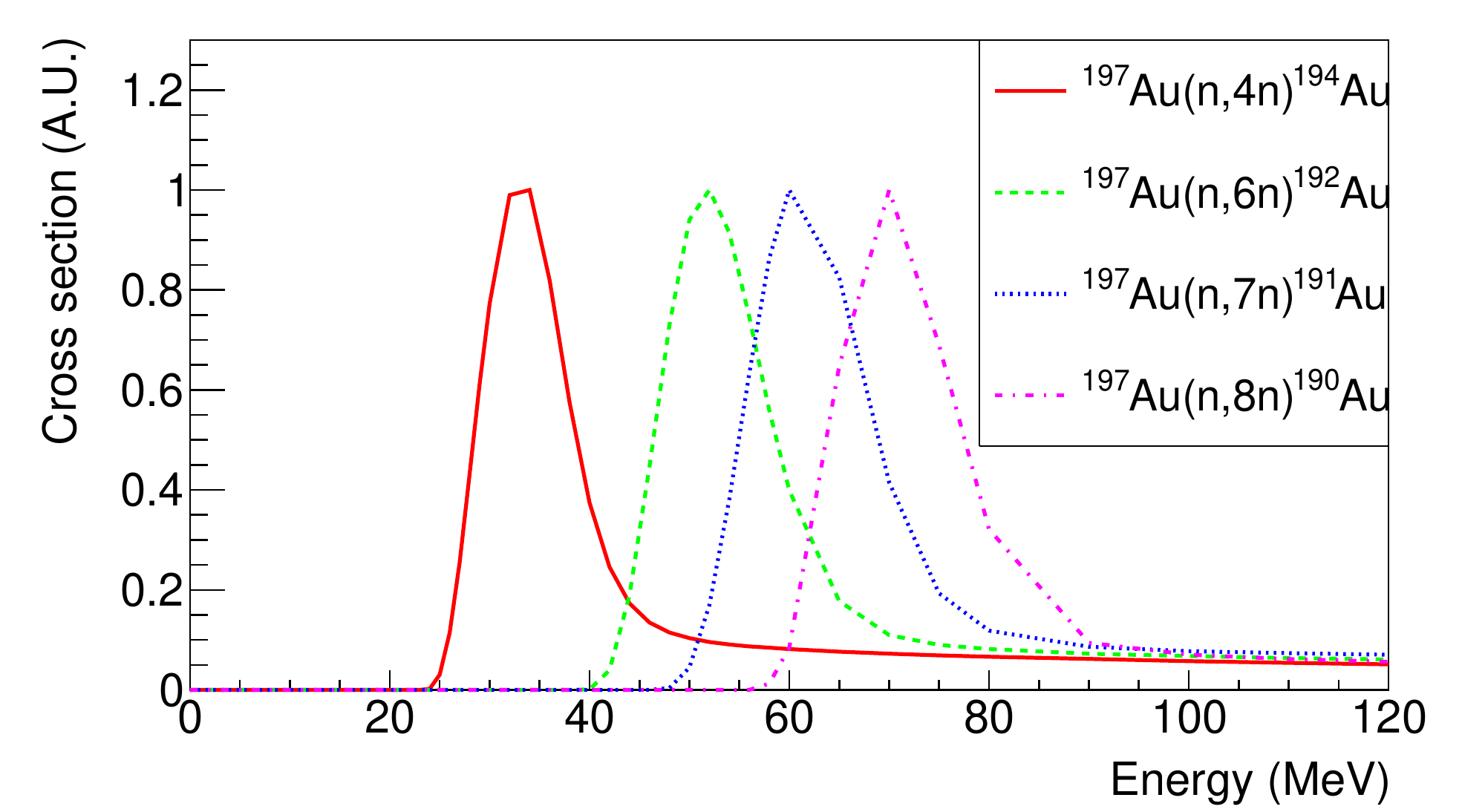}}
\subfloat{\includegraphics[width=0.49\textwidth]{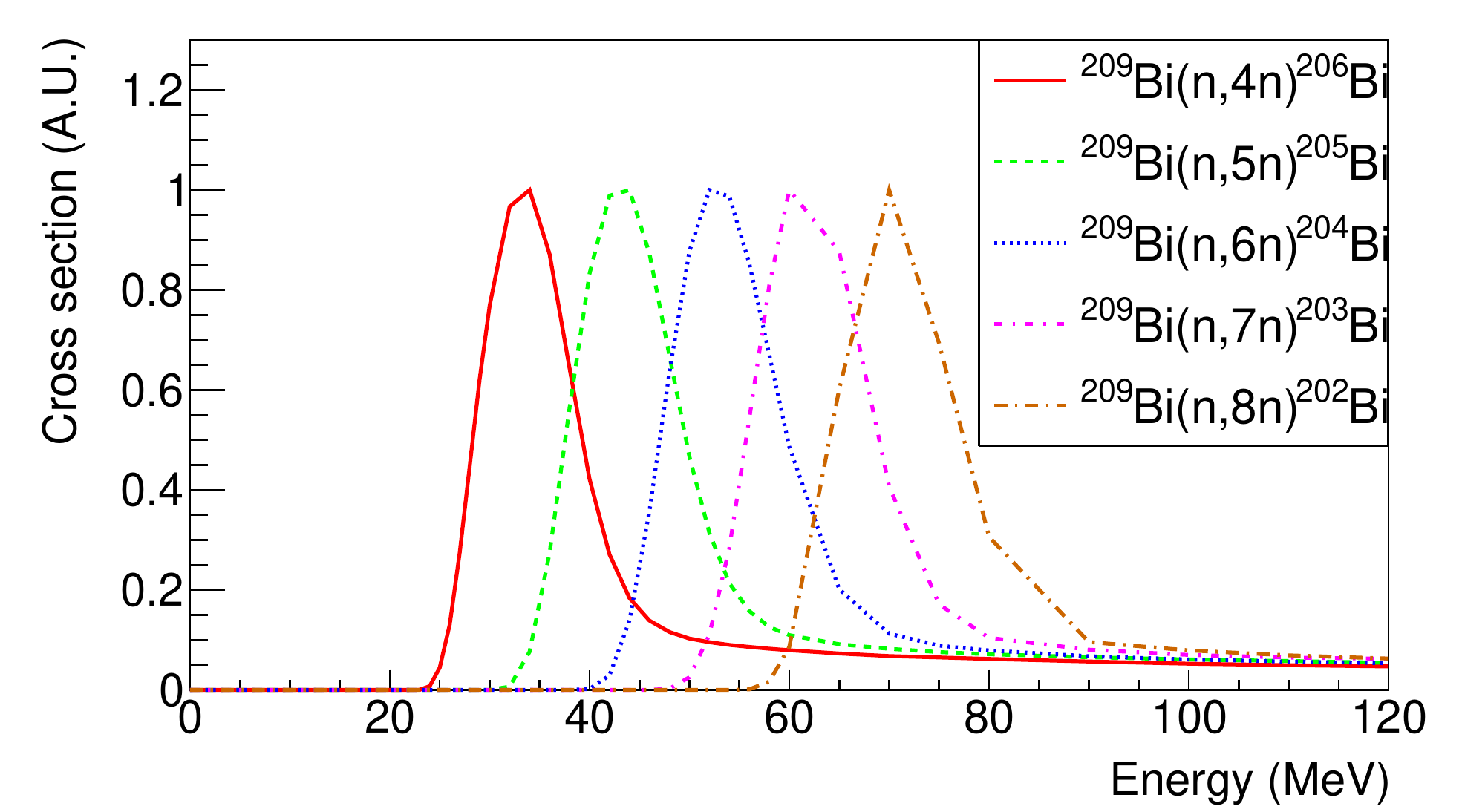}}
\caption{Cross sections of the activation reactions used to unfold the neutron spectrum of ChipIr beamline. For visual purposes, we renormalized each cross section so to make its maximum equal to 1. We show the $(n,\gamma)$ reaction cross sections in the top pads, and the threshold reactions produced by fast neutrons in the central and bottom pads.}
\label{Fig:XSplots}
\end{figure}

To better explain this, we refer to Fig.~\ref{Fig:XSplots}. In these plots, we show the cross sections of the activation reactions used for the analysis presented in this paper. For visual purposes, we renormalized each cross section so to make its maximum equal to 1. In this way, we get an overview of the sensitivity of each reaction to neutrons of different energies. This helps in defining the energy binning for the spectrum unfolding. 

The general recipe for a good choice of binning is based on the following points: first, we must ensure that there is at least one reaction produced in significant quantities by neutrons belonging to each flux group; second, in order to avoid large correlations between flux groups, we must not define more than one group in the energy regions where the cross sections exhibit the same dependence as function of neutron energy. For example, if we tried to split the flux of thermal neutrons in two groups, we would get fully anti-correlated results, because in the thermal region all activation cross sections are proportional to $v^{-1}$ ($v$ being the velocity of the absorbed neutron).

Following this logic, in the region of epithermal and intermediate neutrons we define energy bins that include at least one of the main resonances of the different activation cross sections. In the region of fast neutrons, the binning choice is mainly driven by the different energy thresholds of the observed reactions.

%% file: experiment.tex
According to the methodology explained in the previous section, many different activation reactions must be measured to allow for a successful unfolding of the neutron flux spectrum in a wide energy range.
For this purpose, we irradiated different activation foils with certified purity and elemental composition~\cite{Shieldwerx, GoodFellow}.

The list of the irradiated samples is reported in Tab.~\ref{tab:sample_list}, together with the corresponding weight, thickness and irradiation time. 
For each sample, we have previously evaluated the expected activation rate for the reactions of interest and, depending on the half-life of the isotopes, we tuned the irradiation time to obtain activities in the range that allows to give detectable signals without issues related to radioprotection or excessive dead time during $\gamma$-spectroscopy measurements.

We chose small-sized samples in order to avoid effects related to spatial un-uniformities of the beam. Particularly, all the foils have a diameter of 1.27~cm (0.5~inch) except the Bismuth ones (being squares with 2.5~cm side), to be compared with the ChipIr beam, which is reported to be uniform within a 7$\times$7~cm$^2$ area.
Moreover, in order to minimize self-shielding effects thin foils were used. When self-shielding could not be avoided, a Monte Carlo simulation was performed to calculate the correction to be accounted for the beam attenuation inside the sample. 

\begin{figure}[b!]
\begin{center}
\subfloat[]{\includegraphics[width=0.45\textwidth]{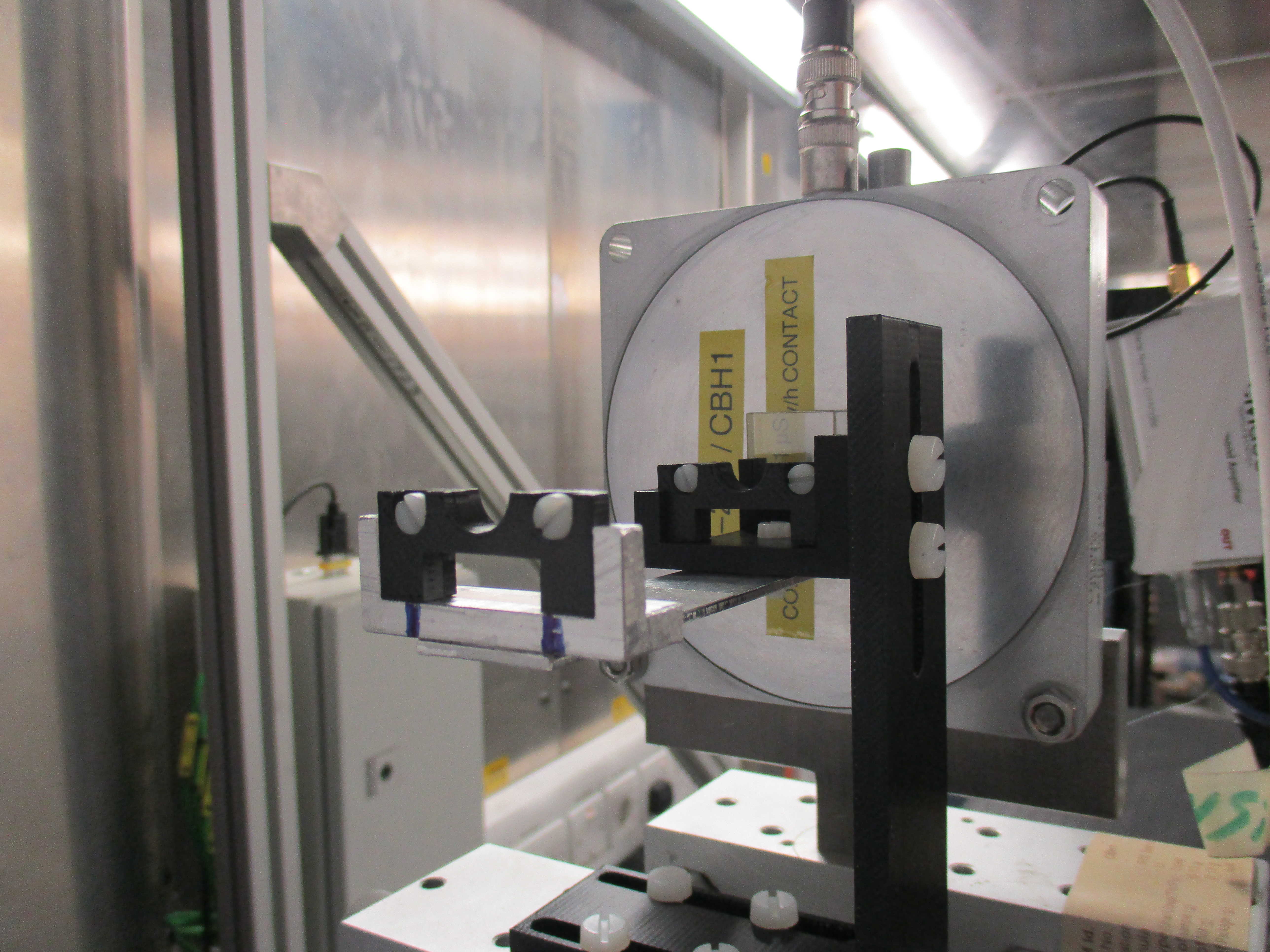}}\qquad
\subfloat[]{\includegraphics[width=0.45\textwidth]{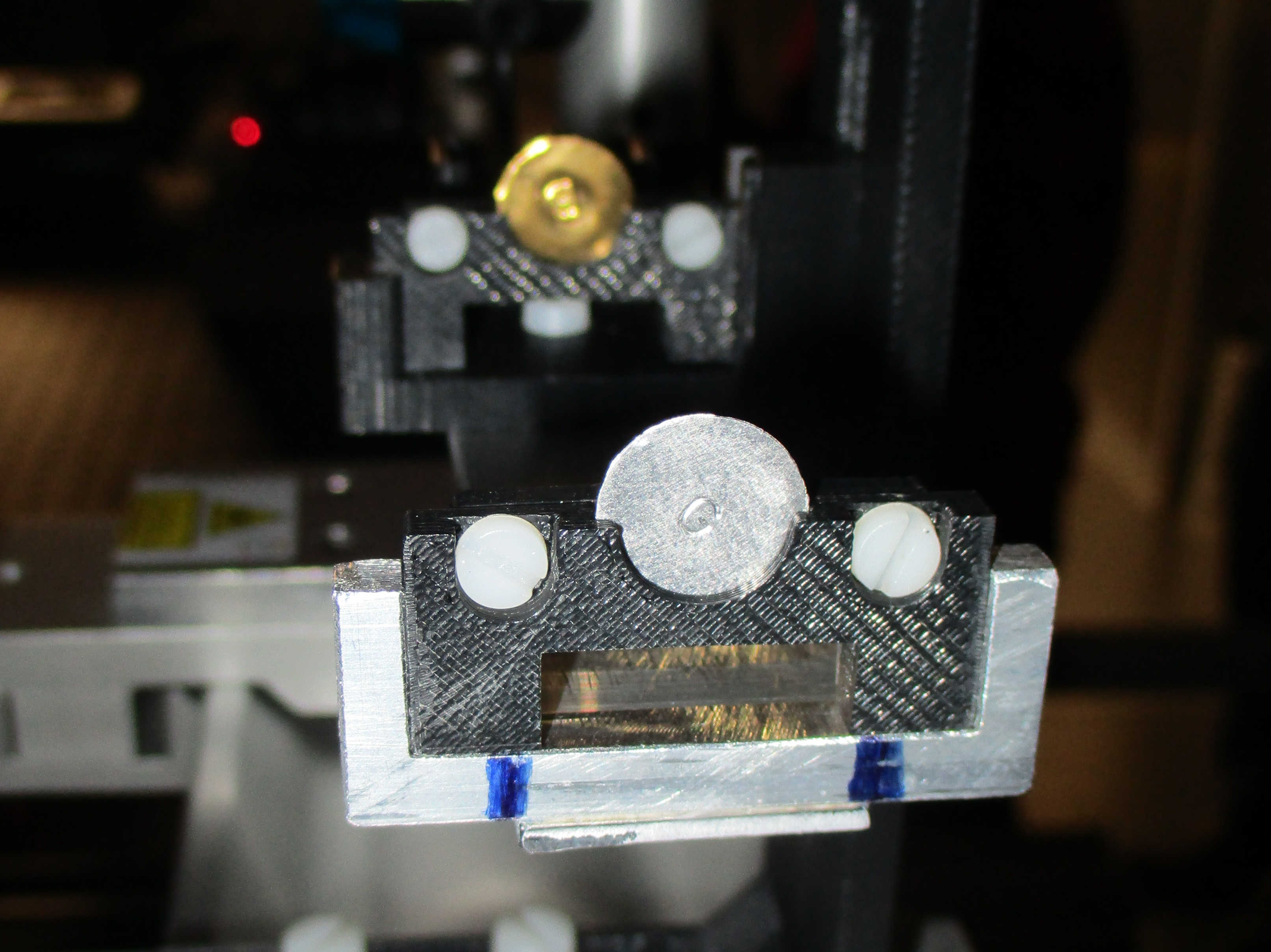}}\\
\subfloat[]{\includegraphics[width=0.45\textwidth]{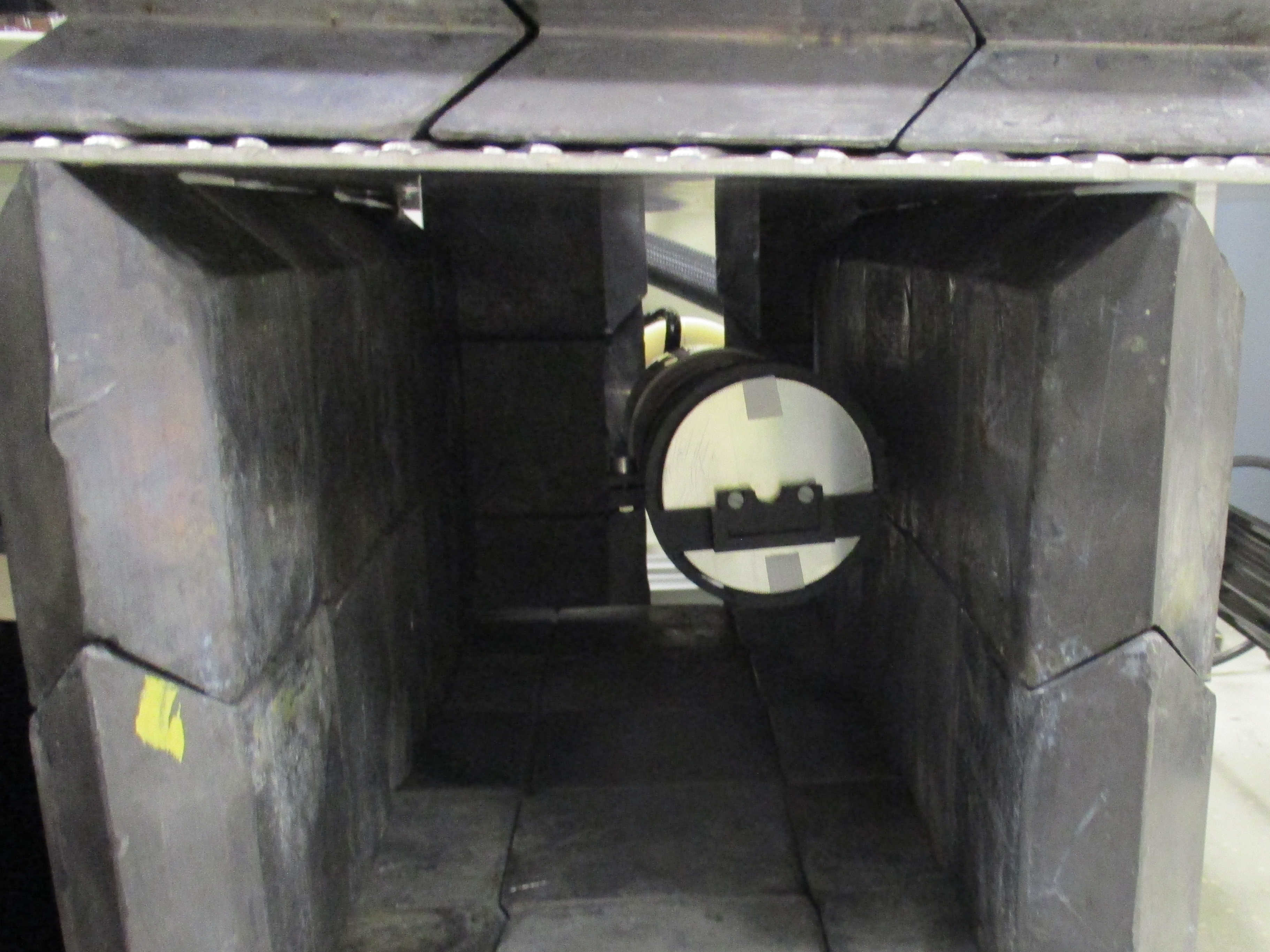}}\qquad
\subfloat[]{\includegraphics[width=0.45\textwidth]{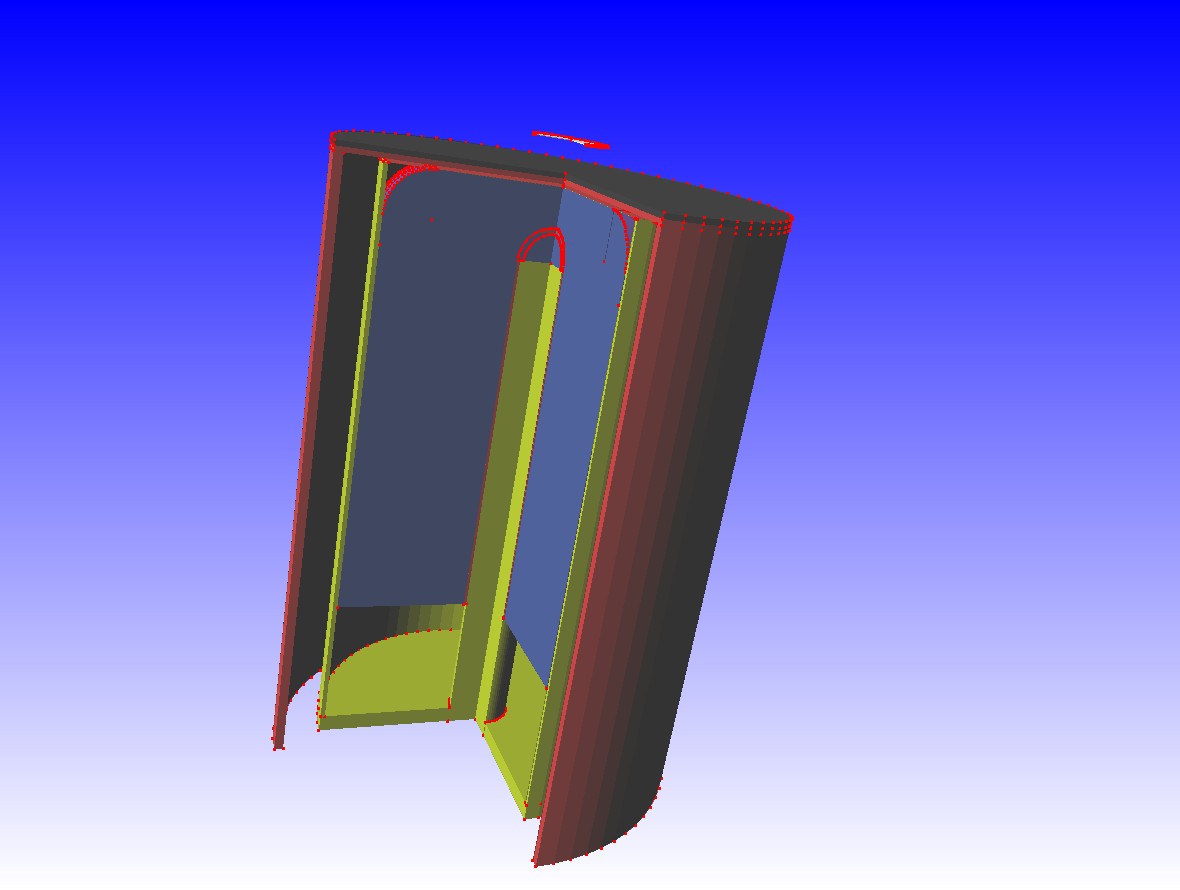}}
\end{center}
\caption{Experimental set-up used for foil irradiations: the sample irradiation holder, placed in the centre of the ChipIr beamline (a) with two samples on it (b). The HPGe detector housed in a lead shielding (c) and its description in the MC simulation model used to evaluate its efficiency (d). }
\label{Fig:experiment}
\end{figure}

\begin{table}[b!]
\footnotesize
\begin{center}
\begin{tabular}{|c|c|c|c|}
\hline
Sample & Weight [g] & Thickness [cm] & $t_{\text{irr}}$ [s]\\
\hline
\hline
Sc 		&		0.0469	& 0.0127  &	5560	\\
Ni 		&		0.2823	&  0.0254 &	51369	\\
MnCu (Mn 83.1\%)	&	0.0488	&  0.00508 &	1675	\\
V 	&	0.0433	& 0.00508 &	979	\\
Zr	&	0.1112	& 0.0127 &	7536	\\
Mo 	&	0.0935	& 0.00762 &	3255	\\
W 	&	0.3191	& 0.0127 &	4515	\\
Au Foil 1	&	0.128	& 0.00508 &	849	\\
AlLu (Lu 5.2\%)	&	0.0333	& 0.01016 &	45024	\\
In 	&	0.1272	& 0.0127 &	9305	\\
Co Foil 1	&	0.0668	& 0.00508 &	9474	\\
AlAu (Au 0.134\%)	&	0.0457	& 0.0127 &	53699	\\
\hline 
Fe 	&	0.1338	&0.0127  &	5560	\\
Bi Foil 1	&	12.579	& 0.2 &	51369	\\
Mg 	&	0.0305	&0.0127  &	7536	\\
Ti 	&	0.1433	& 0.0254 &	45024	\\
Au Foil 2	&	0.1267	& 0.00508&	9305	\\
Bi Foil 2	&	13.127	& 0.2 &	3579	\\
Al 	&	0.0429	&0.0127  &	1622	\\
Co Foil 2	&	0.0639	& 0.00508 &	53699	\\
\hline
\end{tabular}
\end{center}
\caption{List of the irradiated samples. The weight and the thickness of each sample is reported together with the irradiation time, $t_{\text{irr}}$. The samples above (below) the separation line at the center of the table were irradiated in the front (rear) position of the holder. For Au, Co, and Bi, we irradiated two different foils each, changing the irradiation time and/or the position to better detect the different activation reactions.}
\label{tab:sample_list}
\end{table}

An \emph{ad-hoc} holder, made by ABS plastic and aluminum, was used to keep the samples in the centre of the ChipIr neutron beam at a distance of 10.5 meters from the spallation target, see Fig.~\ref{Fig:experiment}~(a--b). 
In designing the holder, great care was devoted in order to reduce the total amount of material around the sample to minimize the neutron flux perturbation due to neutron scattering from holder to samples.
For a more efficient use of the available beam time, the holder allows the simultaneous irradiation of two samples. While the front position is suitable for all reaction types, the rear position can only be used to irradiate foils dedicated to measure threshold reactions. Indeed, the fast neutron spectrum is not significantly modified by the presence of a sample in the front position.

During the irradiation, we monitored the beam status by recording the current of the accelerator and by directly measuring the interaction of neutrons in a diamond detector installed behind the activation foils. This allows to calculate the net irradiation time and to correctly estimate the activation rate in case of beam OFF for some time intervals.

After the neutron irradiation, the sample is placed in front of a fully characterized High Purity Germanium (HPGe) detector to measure the activation spectrum. The detector, also employed for the measurements of the spectral gamma background components at ISIS beamlines~\cite{FestaIMAT,PIETROPAOLO2006826,C4JA00183D}, is a coaxial HPGe (GMX series by ORTEC) with carbon endcap and relative efficiency of 40\%, and was housed in a lead shielding to limit the background Fig.~\ref{Fig:experiment}~c.

In most cases, especially when several isotopes with different half-lives were produced, we measured the samples a few times (usually, two or three). The first measurements after irradiation are needed to detect the $\gamma$-rays of isotopes with relatively short half-life, while measurements after the decay of short lived isotopes allow higher sensitivity to isotopes with lower activity and longer decay time.

%% file: results.tex
The data analysis to obtain the measurement of the neutron flux and its spectrum consists of two main steps. First, we analyze the measured spectra of the $\gamma$-rays emitted by the irradiated samples to evaluate the activity of the produced radioisotopes and, from this, the corresponding activation rates. Second, we use the data of the measured activation rates to unfold the neutron spectrum according to the method described in Sect.~\ref{sec:method}.

\subsection{Analysis of gamma spectroscopy measurements}

Given the number of counts ($n_{\text{counts}}$) in a $\gamma$-line, it is possible to estimate the number of decays ($n_{\text{dec}}$) occurred during the measurement --and thus the absolute activity-- if the intensity of the $\gamma$-line ($I_{\gamma}$) and the detection efficiency ($\epsilon$) are known.
\[n_{\text{counts}} = I_{\gamma}\,\epsilon\,n_{\text{dec}} \]
The product $I_{\gamma}\,\epsilon$ represents, for each $\gamma$-line, the probability that, given a decay in the sample, an event is recorded in the full-energy peak.
To evaluate this probability for each $\gamma$-line, we exploit Monte Carlo (MC) simulations based on \textsc{Geant4} toolkit~\cite{GEANT4}. Simulations generate the decay of different radioisotopes and record the energy depositions in the detector. The model includes detailed geometries and materials of samples and detector (Fig.~\ref{Fig:experiment}~d). This MC tool provides a simulated energy spectrum, after generating a given number of decays ($n^{\textsc{mc}}_{\text{dec}}$) in the source volume.
In this way we can evaluate the detection probability as
\[ I_{\gamma}\,\epsilon = \dfrac{n^{\textsc{mc}}_{\text{counts}}}{n^{\textsc{mc}}_{\text{dec}}} \]
where $n^{\textsc{mc}}_{\text{counts}}$ is the number of counts in the full-energy peak of the simulated spectrum.

Measurements with certified activity calibration have been performed to validate the accuracy of the MC model. The compatibility between measurements and simulations remain within the 5\% for all $\gamma$-lines. This result is in agreement with what has been obtained in other activation analyses~\cite{AbsoluteFlux, FluxDistribution} and provides an estimate of the systematic uncertainty affecting the absolute activity measurements with this method.\\

\begin{figure}[!htb]
\begin{center}
\subfloat[]{\includegraphics[width=0.49\textwidth]{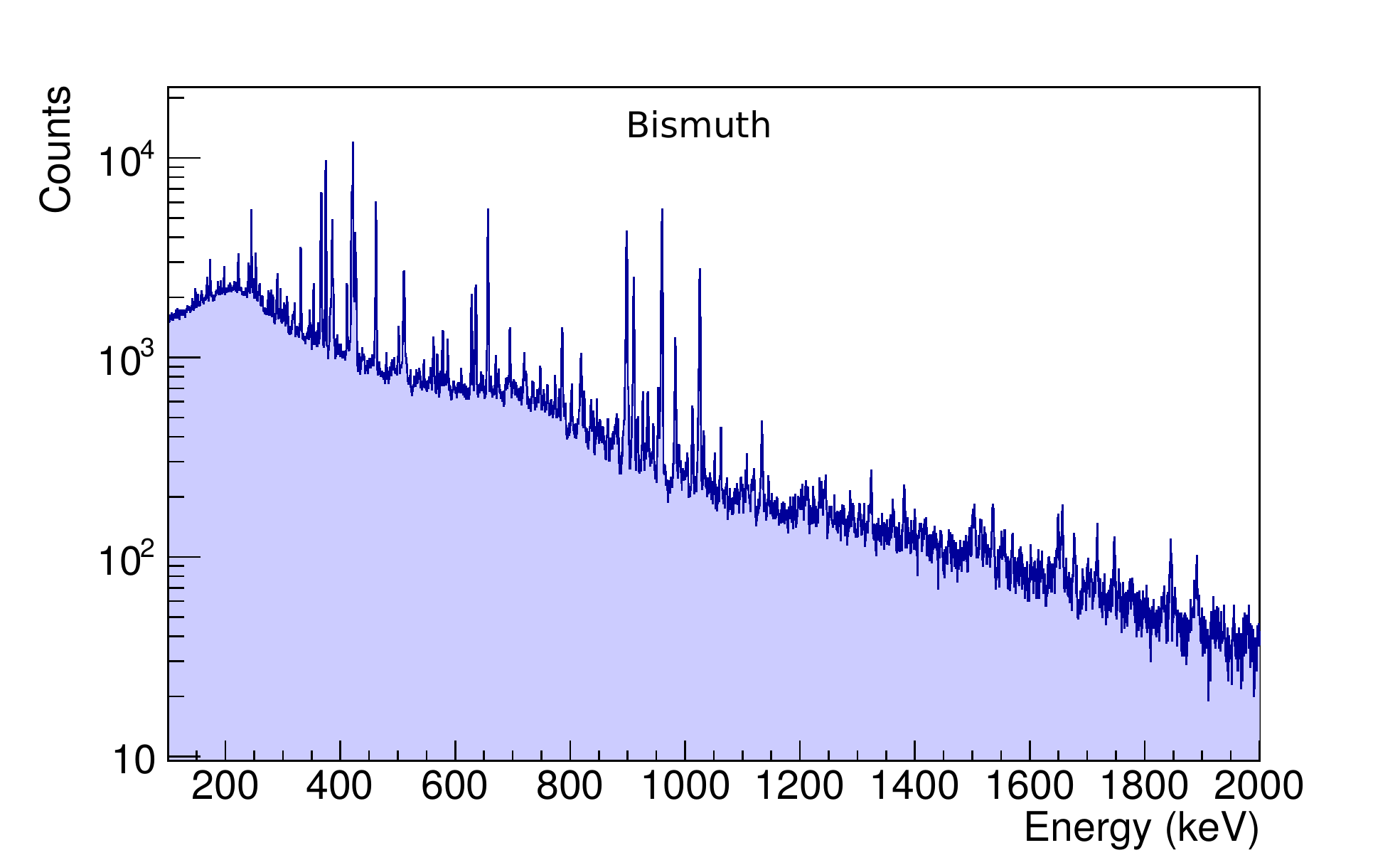}}
\subfloat[]{\includegraphics[width=0.49\textwidth]{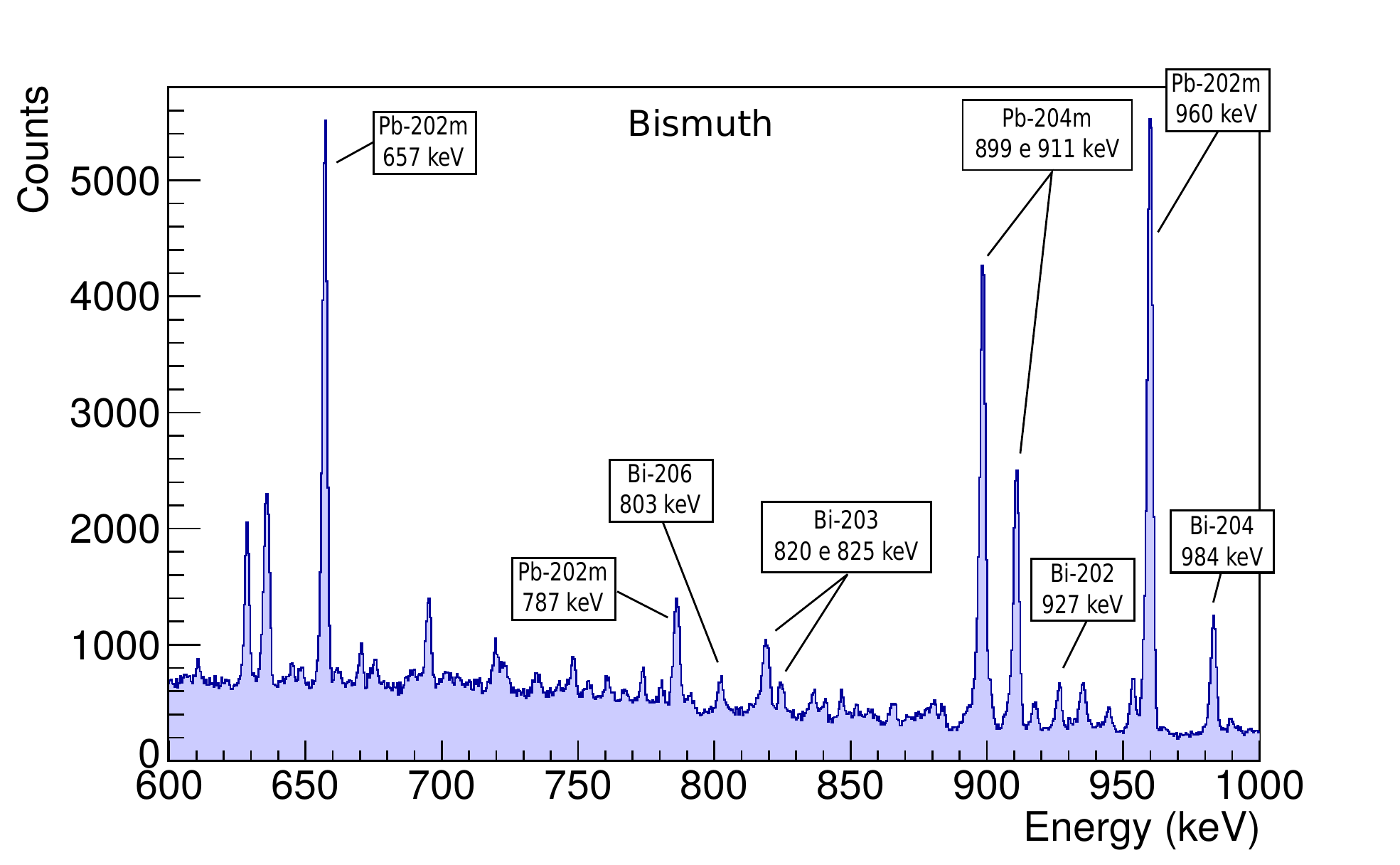}}\\
\subfloat[]{\includegraphics[width=0.49\textwidth]{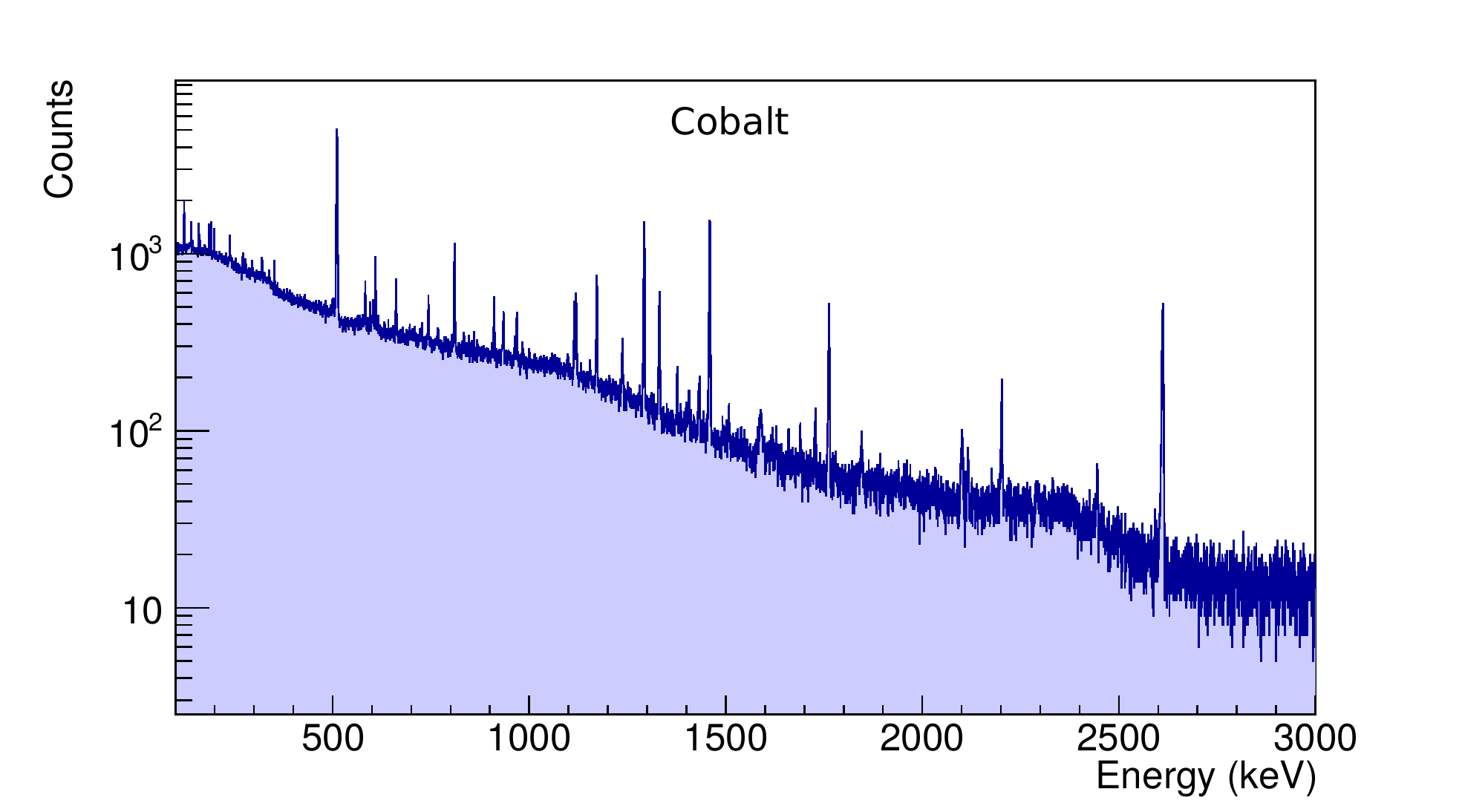}}
\subfloat[]{\includegraphics[width=0.49\textwidth]{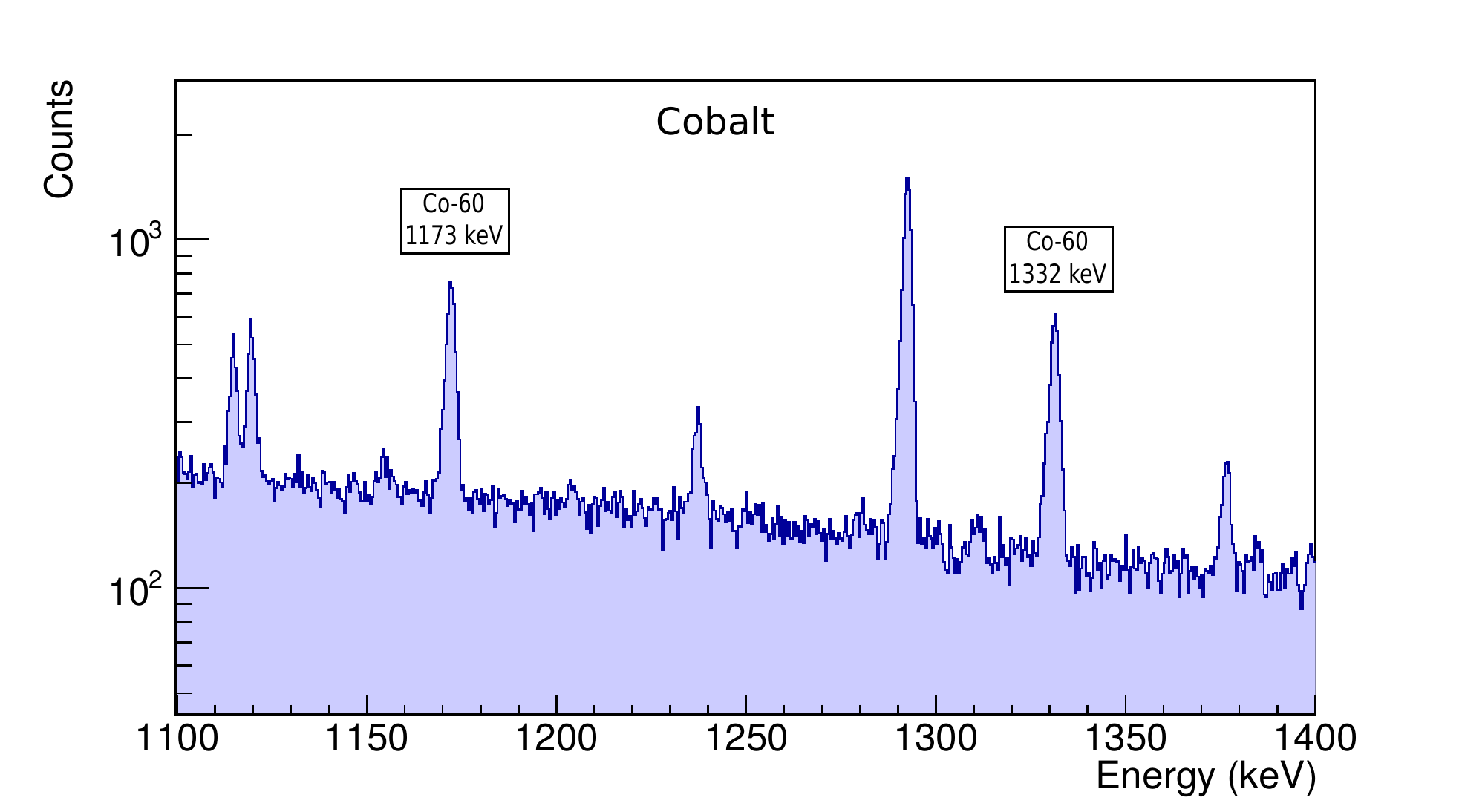}}\\
\end{center}
\caption{Examples of experimental $\gamma$-spectra measured by the HPGe after the activation of bismuth and cobalt foils. }
\label{Fig:GammaSpectra}
\end{figure}

For each experimental spectrum, we estimate the net number of counts in the peaks, subtracting the background events. Particularly, we fit the $\gamma$-lines with Gaussian functions and the background with second degree polynomials. The same fit procedure is applied to both the experimental and the MC spectra. \\

The analysis of the experimental spectra begins with the identification of the isotopes that emit the observed $\gamma$-lines. This is done using the tabulated energies for each isotope.
Fig.~\ref{Fig:GammaSpectra} shows two examples of spectra recorded with the HPGe detector after the activation of bismuth and cobalt foils. The plots on the right are expanded views of two portions of these spectra, with labels indicating the isotopes that produce some of the $\gamma$-lines.

As shown in these examples, in some cases the spectra exhibit many $\gamma$-lines that can overlap. A failure to identify an overlap of peaks generated by two different isotopes would lead to a wrong estimate of the corresponding activities. To avoid this kind of mistake, we cross check the lists of the $\gamma$-rays emitted by the isotopes that are expected to be produced in the foil. Moreover, we exploit MC simulations to carry out a comparative analysis of the relative intensities of the peaks emitted by the same isotope. Indeed, in the absence of overlaps with $\gamma$-lines emitted by other isotopes, we expect to observe the same relative intensities of peaks in the experimental and MC spectra. 
When we find an overlap of peaks that cannot be separated, we finally decide to reject their data, in order to avoid systematic errors.

\subsection{Activation Rate Evaluation}

For each observed $\gamma$-line, we evaluate the absolute activity of the emitting isotope at the beginning of the spectroscopy measurement:
\[A = \dfrac{n_{\text{dec}} \lambda} {\left( 1-e^{-\lambda t_{\text{meas}} } \right) } = \dfrac{n_{\text{counts}} \lambda}{\left( 1-e^{-\lambda t_{\text{meas}} } \right) } I_{\gamma}\,\epsilon \]
where $t_{\text{meas}}$ is the measurement livetime. 

\begin{figure}[b!]
\begin{center}
\subfloat[]{\includegraphics[width=0.49\textwidth]{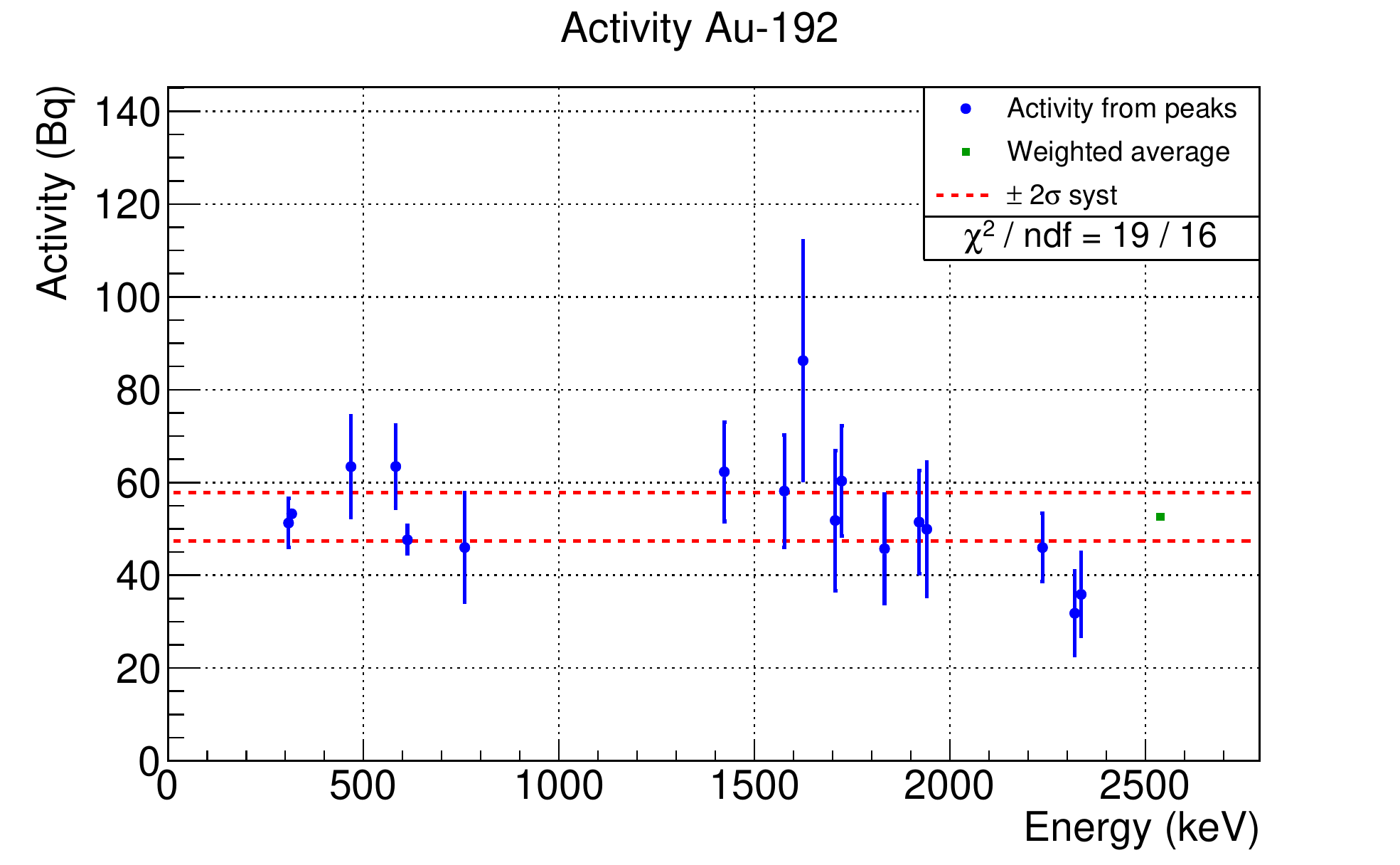}}
\subfloat[]{\includegraphics[width=0.49\textwidth]{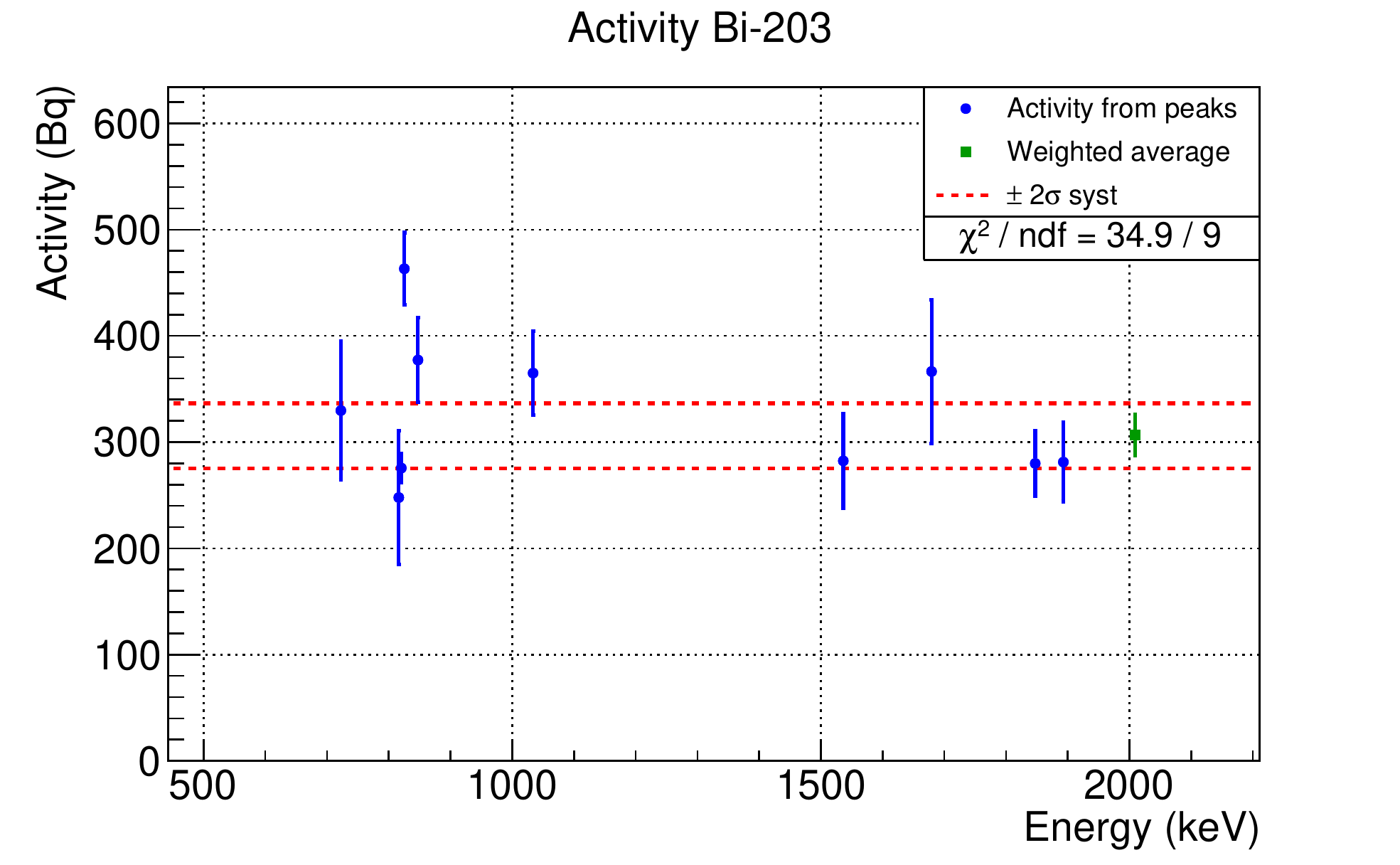}}
\end{center}
\caption{Activity of $^{192}$Au and $^{203}$Bi measured with $\gamma$-lines of different energies. The round (blue) points are the activities calculated from the data of each peak, separately. The squared (green) point is the weighted average of the activities. The dotted (red) horizontal lines are used to represent the $\pm2\sigma$ global systematic uncertainty due to efficiency evaluation with MC simulations.}
\label{Fig:Activity}
\end{figure}

When more than one $\gamma$-line is available for the same radioisotope, we expect to obtain activity evaluations compatible within the statistical uncertainty. The plots in Fig.~\ref{Fig:Activity} show two examples in which we observed several $\gamma$-lines emitted by the same isotopes: $^{192}$Au and $^{203}$Bi, respectively. 
In general, as shown in these examples, there is good agreement between the activity evaluations from the data of each peak, proving that our methodology is robust.
To get the best estimate of the activity, we compute the weighted average of the results obtained from the data of each peak. 
Then, to quantify the agreement between the activities calculated from each peak, we perform a $\chi^2$ test. 
In some cases, such as that of $^{203}$Bi in Fig.~\ref{Fig:Activity}, the ratio between the $\chi^2$ and the number of degrees of freedom (ndf) is greater than 1, meaning that there is some additional uncertainty affecting the data. Possible source of error can be related, for example, to the evaluation of $n_{\text{counts}}$ through the fit or to the MC simulations of the isotope decays. In those cases, we increase the uncertainty associated to the weighted average by a factor given by $\sqrt{\chi^2 / \text{ndf}}$, according to the prescription of the Particle Data Group~\cite{PDG2016}.\\

\begin{table}[b!]
\begin{center}
\begin{tabular}{|lc|lc|}
\hline
Reaction & Act.Rate/m (s$^{-1}$g$^{-1}$) & Reaction & Act.Rate/m (s$^{-1}$g$^{-1}$)\\
\hline
$^{45}$Sc(n,$\gamma$)	&	$(1.33\pm0.07)\times10^{5}$	&	$^{59}$Co(n,2n)$^{58}$Co	&	$(4.99\pm0.29)\times10^{3}$	\\
$^{51}$V(n,$\gamma$)	&	$(1.73\pm0.12)\times10^{4}$	&	$^{59}$Co(n,3n)$^{57}$Co	&	$(2.98\pm0.18)\times10^{3}$	\\
$^{55}$Mn(n,$\gamma$)	&	$(5.23\pm0.28)\times10^{4}$	&	$^{59}$Co(n,$\alpha$)$^{56}$Mn	&	$(4.91\pm0.27)\times10^{2}$	\\
$^{59}$Co(n,$\gamma$)	&	$(1.72\pm0.10)\times10^{5}$	&	$^{nat}$Ni(n,*)$^{57}$Ni	&	$(1.03\pm0.05)\times10^{3}$	\\
$^{64}$Ni(n,$\gamma$)	&	$(6.27\pm0.73)\times10^{1}$	&	$^{nat}$Ni(n,*)$^{61}$Co	&	$(1.48\pm0.08)\times10^{2}$	\\
$^{96}$Zr(n,$\gamma$)	&	$(1.08\pm0.09)\times10^{2}$	&	$^{nat}$In(n,*)$^{113m}$In	&	$(2.66\pm0.19)\times10^{2}$	\\
$^{100}$Mo(n,$\gamma$)	&	$(2.31\pm0.15)\times10^{2}$	&	$^{115}$In(n,n')$^{115m}$In	&	$(9.06\pm0.47)\times10^{2}$	\\
$^{nat}$In(n,*)$^{114m}$In	&	$(8.73\pm0.49)\times10^{3}$	&	$^{197}$Au(n,4n)$^{194}$Au	&	$(2.18\pm0.11)\times10^{3}$	\\
$^{115}$In(n,$\gamma$)$^{116m}$In	&	$(4.80\pm0.24)\times10^{5}$	&	$^{197}$Au(n,6n)$^{192}$Au	&	$(1.65\pm0.08)\times10^{3}$	\\
$^{176}$Lu(n,$\gamma$)	&	$(1.25\pm0.08)\times10^{5}$	&	$^{197}$Au(n,7n)$^{191}$Au	&	$(1.18\pm0.07)\times10^{3}$	\\
$^{186}$W(n,$\gamma$)	&	$(3.23\pm0.17)\times10^{4}$	&	$^{197}$Au(n,8n)$^{190}$Au	&	$(5.16\pm0.33)\times10^{2}$	\\
\textit{Au}: $^{197}$Au(n,$\gamma$)	&	$(1.82\pm0.12)\times10^{5}$	&	$^{209}$Bi(n,4n)$^{206}$Bi	&	$(2.15\pm0.11)\times10^{3}$	\\
\textit{Al-Au}: $^{197}$Au(n,$\gamma$)	&	$(4.40\pm0.24)\times10^{5}$	&	$^{209}$Bi(n,5n)$^{205}$Bi	&	$(1.91\pm0.11)\times10^{3}$	\\
$^{nat}$Mg(n,*)$^{24}$Na	&	$(2.43\pm0.15)\times10^{3}$	&	$^{209}$Bi(n,6n)$^{204}$Bi	&	$(1.38\pm0.08)\times10^{3}$	\\
$^{27}$Al(n,p)$^{27}$Mg	&	$(1.23\pm0.10)\times10^{3}$	&	$^{209}$Bi(n,7n)$^{203}$Bi	&	$(1.17\pm0.07)\times10^{3}$	\\
$^{27}$Al(n,$\alpha$)$^{24}$Na	&	$(2.22\pm0.11)\times10^{3}$	&	$^{209}$Bi(n,8n)$^{202}$Bi	&	$(9.27\pm0.66)\times10^{2}$	\\
$^{nat}$Fe(n,*)$^{56}$Mn	&	$(6.94\pm0.42)\times10^{2}$	&		&		\\
\hline
\end{tabular}
\end{center}
\caption{Complete list of the measured activation rates per unit mass on ChipIr. The $^{197}$Au(n,$\gamma$) reaction was measured with both \textit{pure Au} and \textit{Al-Au alloy} foils, obtaining very different activation rates due to the non negligible self-shielding effect in the \textit{pure Au} sample.}
\label{tab:ActRate}
\end{table}

As described in Sect.~\ref{sec:experiment}, almost all the activated foils were measured more than once. In many cases, we observe the peaks of the same isotopes in spectroscopy measurements done at different times. To compare the activities evaluated from these measurements, we choose a common reference time and we calculate:
\[A_0 = \dfrac{A}{e^{-\lambda t_{\text{wait}}}}\]
where $t_{\text{wait}}$ is the time elapsed between the end of the irradiation and the beginning of the measurement.
After checking the statistical compatibility of the resulting activities, we calculate their weighted average.

Finally, we use our best estimates of $A_0$ to calculate the activation rates of the isotopes:
\[R = \dfrac{A_0}{1-e^{-\lambda t_{\text{irr}}}}\]
where $t_{\text{irr}}$ is the irradiation time.

In some cases, we irradiated various foils containing the same elements and we observed the same activation reactions.
This allows us to compare the results obtained from independent measurements performed in the same experimental conditions. For a direct comparison of the results, we divide the activation rate by the mass of the target element in the foil. Again, we obtain a very good agreement of the results, confirming the reliability of the experimental measurements and of the data analysis.    

The list of the observed reactions with the corresponding activation rates per unit mass is reported in Tab.~\ref{tab:ActRate}. The uncertainties presented here include the aforementioned 5\% systematic uncertainty related to the absolute estimate of activity based on the MC model of gamma spectroscopy measurements.

\subsection{Neutron flux spectrum unfolding}

In the second step of the analysis, we perform the neutron flux spectrum unfolding with the method presented in Sect.~\ref{sec:method}, using the experimental data in Tab.~\ref{tab:ActRate}. 

\begin{table}[b!]
\begin{center}
\begin{tabular}{|c|ll|ll|ll|}
\hline
Group & \multicolumn{2}{|c|}{Energy range (MeV)} & \multicolumn{2}{|c|}{Reference unfolding} & \multicolumn{2}{|c|}{Minimal unfolding} \\
number  & Min & Max &	\multicolumn{2}{|c|}{Flux ($10^5$cm$^{-2}$s$^{-1}$)}	&	\multicolumn{2}{|c|}{Flux ($10^5$cm$^{-2}$s$^{-1}$)}		\\
\hline
1 & $10^{-9}$		&	$4\times10^{-8}$	&	$1.12\pm0.16$	&	(15\%)	&	$1.13\pm0.15$	&	(13\%)	\\
2 & $4\times10^{-8}$	&	$5\times10^{-7}$	&	$2.90\pm0.38$	&	(13\%)	&	$2.90\pm0.33$	&	(11\%)	\\
3 & $5\times10^{-7}$	&	$3\times10^{-6}$	&	$1.11\pm0.16$	&	(14\%)	&	$1.10\pm0.14$	&	(13\%)	\\
4 & $3\times10^{-6}$	&	$10^{-5}$		&	$0.77\pm0.06$	&	(7\%	)	&	$0.77\pm0.06$	&	(7\%	)	\\
5 & $10^{-5}$		&	$10^{-4}$		&	$2.30\pm0.23$	&	(10\%)	&	$2.27\pm0.19$	&	(8\%	)	\\
6 & $10^{-4}$		&	$10^{-3}$		&	$2.46\pm0.22$	&	(9\%	)	&	$2.46\pm0.21$	&	(9\%	)	\\
7 & $10^{-3}$		&	1				&	$4.15\pm1.58$	&	(38\%)	&	$4.10\pm1.45$	&	(35\%)	\\
8 & 1				&	10				&	$3.88\pm0.95$	&	(24\%)	&	$3.62\pm0.33$	&	(9\%	)	\\
9 & 10				&	20				&	$3.16\pm0.81$	&	(26\%)	&	$3.31\pm0.26$	&	(8\%	) \\
10 & 20				&	40				&	$4.76\pm1.31$	&	(28\%)	&	$5.09\pm0.31$	&	(6\%	)	\\
11 & 40				&	60				&	$4.93\pm1.97$	&	(40\%)	&	$5.46\pm0.33$	&	(6\%	)	\\
12 & 60				&	100				&	$7.03\pm1.35$	&	(19\%)	&	$6.08\pm0.23$	&	(4\%	)	\\
13 & 100			&	700				&	$34.2\pm6.6$		&	(19\%)	&	$29.6\pm1.1$		&	(4\%	)	\\
\hline
 	&		&	Total			&	$72.8\pm6.8$		&	(9\%)	&	$67.8\pm1.8$		&	(3\%)	\\
\hline
\end{tabular}
\end{center}
\caption{Results of the neutron flux intensity in the different energy groups used to unfold the spectrum. The uncertainties are only statistical ones (we report the corresponding relative uncertainties in brackets). In the second column we show the results of the \textit{reference} unfolding, where cross section uncertainties are propagated to the results. In the third column we report the results of the \textit{minimal} unfolding, that does not include cross section uncertainties.}
\label{tab:Results}
\end{table}

Using the criteria discussed above, we identified a good choice for the energy binning (see first column of Tab.~\ref{tab:Results}). 
We define 13 flux groups: 6 groups in the region below 1~keV, 6 groups above 1~MeV and one group to cover the range between 1~keV and 1~MeV. The latter flux group does not have strong experimental signatures, like large resonances or thresholds of the cross sections, that would allow to define more flux groups.

A particular case is the range from 100~MeV to 700~MeV (corresponding to the last bin). Cross section data available in literature for the activation reactions with thresholds above 100~MeV are extremely poor and affected by large uncertainties. 
However, according to independent measurements and simulations reported in literature~\cite{PlattSpectra}, we know that the neutron flux generated by a spallation source like ISIS is characterized by a decreasing spectrum above $\sim100$~MeV. 
This allows to set specific constraints to the flux intensity in the last bin. Therefore, the last group is not a free parameter of the analysis, but it is constrained to have an intensity such that the spectrum is connected with continuity at 100~MeV and then decreases.\\

In order to unfold the neutron spectrum, as explained in Sect.~\ref{sec:method}, we must choose \textit{intra-group} spectrum shapes to calculate the effective cross sections in each energy bin.
As a starting point, we use a spectrum shape proportional to $E^{-1}$ for all groups with two exceptions: the first bin and the bins in the range [1--100]~MeV, for which we use constant spectrum shapes.
A detailed analysis about the systematic uncertainties related to the choice of the \textit{intra-group} spectrum shapes and of the constraint for the last group of flux is presented in Sect.~\ref{sec:syst}.\\

First of all, we present the results of what we call \textit{reference} unfolding (in Tab.~\ref{tab:Results}, second column). 
This is based on a statistical model that includes the uncertainties of the cross sections, so they are propagated to the flux results. In this model, the cross sections are free to vary within their uncertainties to better fit the activation rate data when combined with the evaluations of the flux groups.
When needed, self-shielding correction factors are applied to the effective cross sections. These factors are estimated by comparing the results of MCNP~\cite{mcnp} simulations in which the density of the foils is varied from the real value to an extremely small value, for which self-shielding is negligible.

\begin{figure}[b!]
\begin{center}
\includegraphics[width=0.9\textwidth]{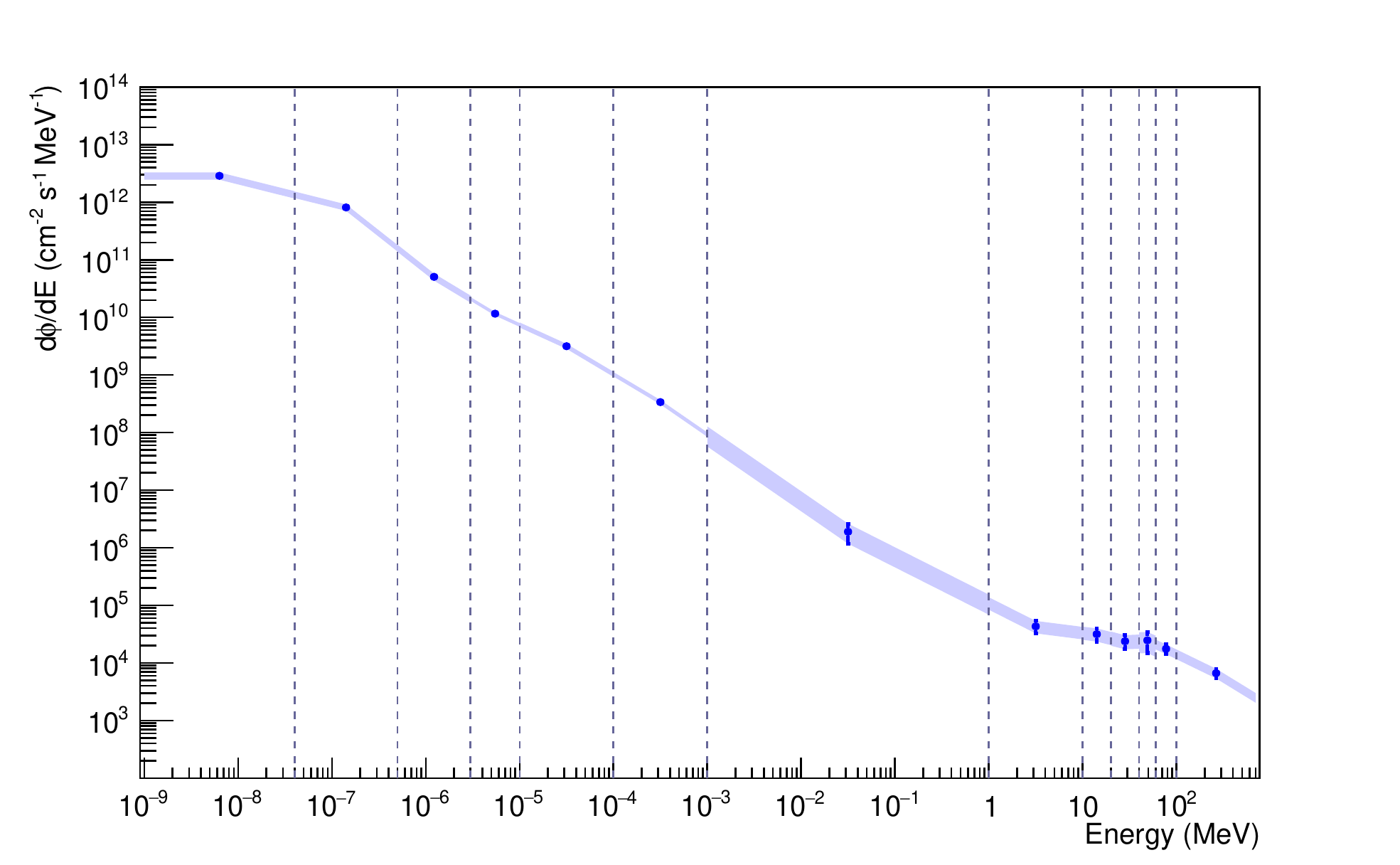}
\end{center}
\caption{The ChipIr neutron flux energy spectrum resulting from the reference unfolding.}
\label{Fig:UnfoldedSpectrum}
\end{figure}

\begin{figure}[t!]
\begin{center}
\includegraphics[width=0.5\textwidth]{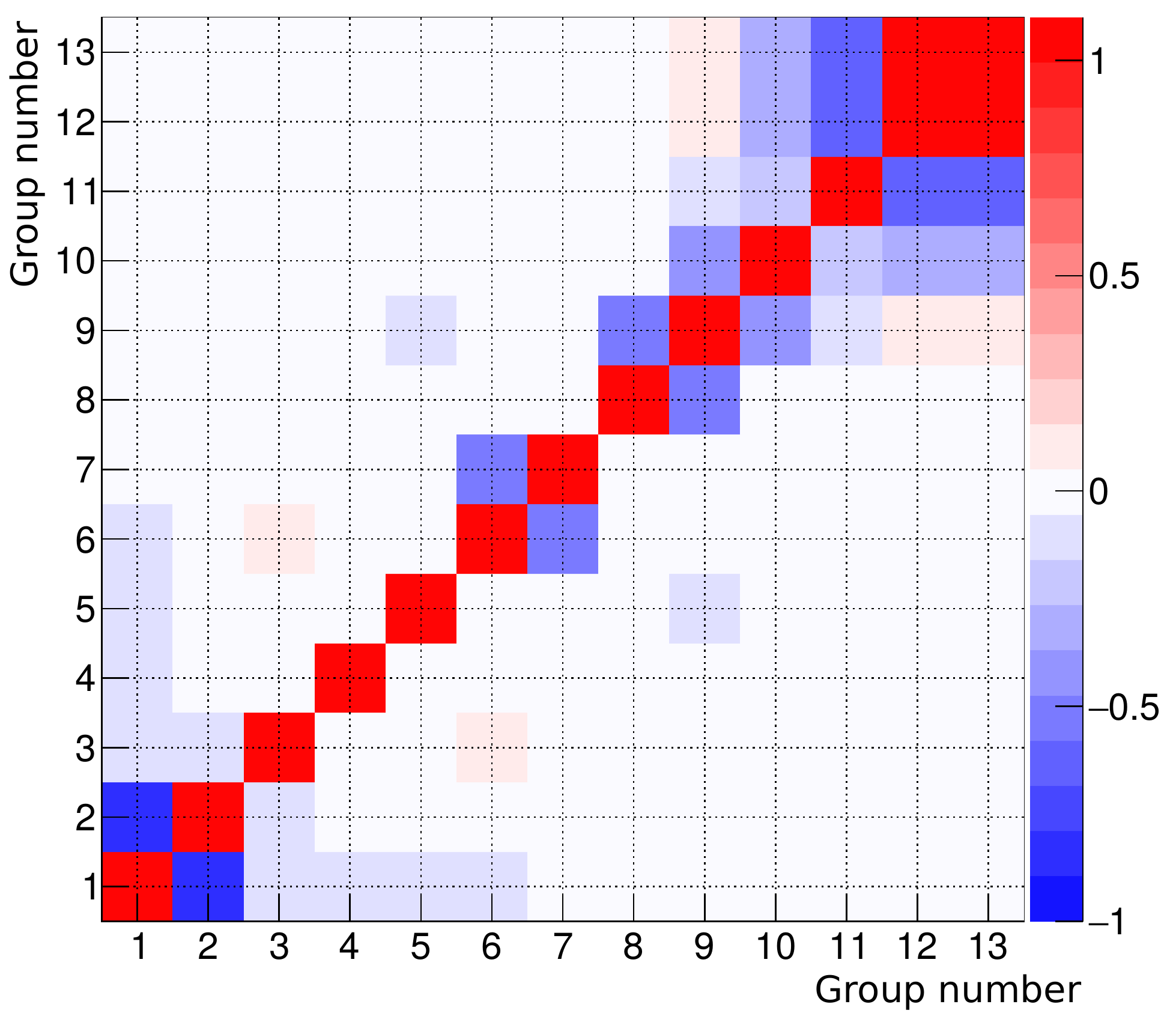}
\end{center}
\caption{Matrix of the correlations between energy groups for the reference unfolding result.}
\label{Fig:CorrMatrix}
\end{figure}

In Fig.~\ref{Fig:UnfoldedSpectrum} we present the unfolded neutron flux spectrum. In this plot, the points are the $d\phi/dE$ values calculated at the center of each bin using the results of the \textit{reference} unfolding, while the stripe that connects the points represents the statistical uncertainty affecting the flux results in the bins used to unfold the neutron spectrum.

The correlation matrix of the results obtained for the group of fluxes is shown in Fig.~\ref{Fig:CorrMatrix}. 
For the most part, the results are not correlated or only weakly correlated. 
However in some cases, there are pairs of groups that exhibit negative correlation. In particular, we observe anti-correlation for the following pairs of adjacent groups: 1--2, 6--7, 8--9, 9--10, and 11--12. 
This feature is quite usual when spectra are unfolded with this technique; the sum of the flux in two adjacent groups is determined with better precision than the flux in the two separate groups. 
The analysis of the correlations shows that the energy binning we chose is a good compromise that allows to obtain, at the same time, a detailed description of the neutron spectrum and limited correlations between the groups.

The results obtained with the \textit{reference} unfolding feature a good precision in the thermal and epithermal range up to 1~keV, while are affected by higher uncertainty in the region of fast neutrons. 
This is mainly due to the fact that the cross sections of the threshold reactions used to unfold the spectrum of fast neutrons are affected by larger uncertainties than the cross sections of the (n,$\gamma$) reactions.

\subsection{Impact of cross section uncertainties}

\begin{figure}[!htb]
\begin{center}
\subfloat{\includegraphics[width=0.7\textwidth]{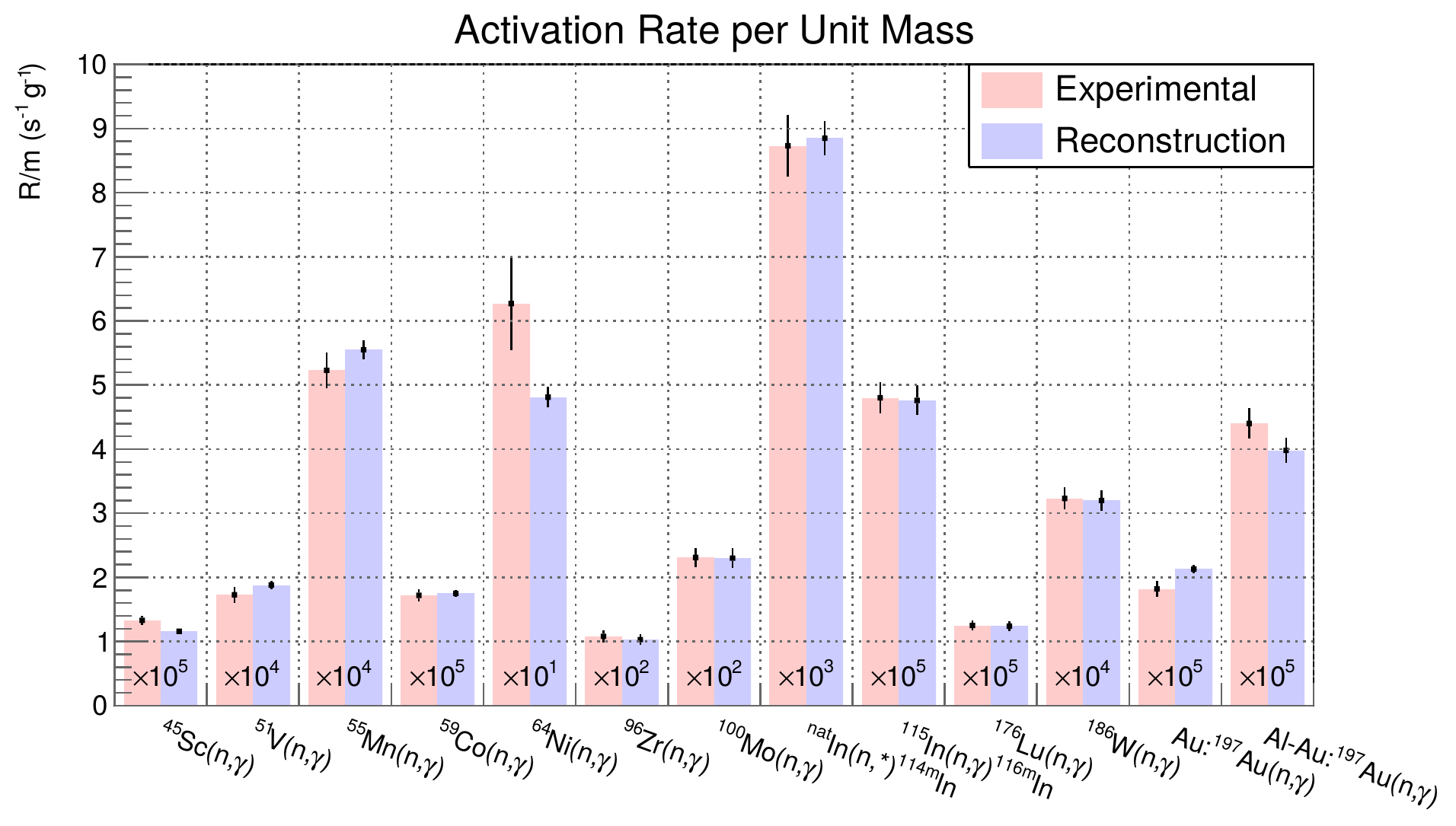}}\\
\subfloat{\includegraphics[width=0.7\textwidth]{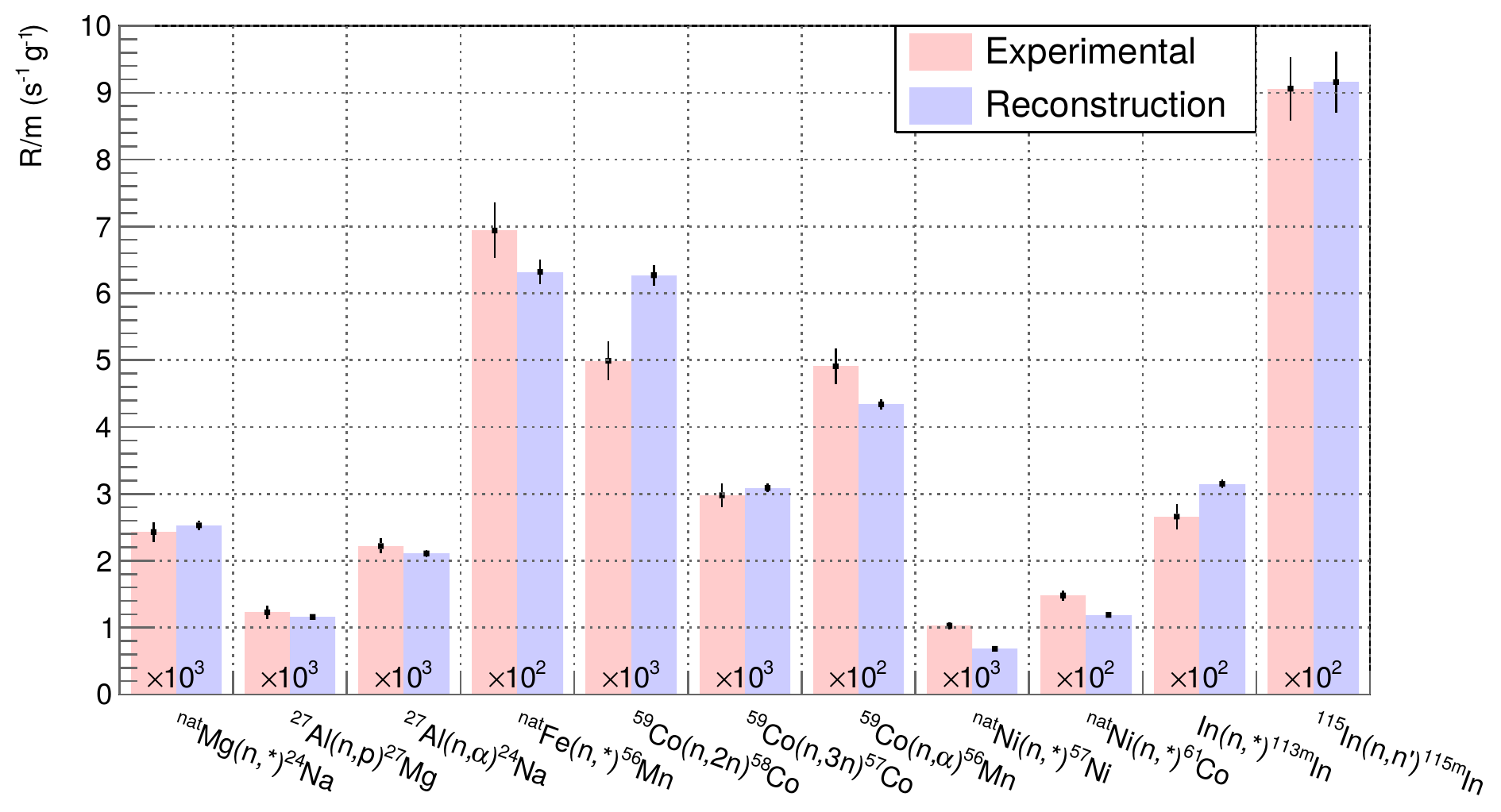}}\\
\subfloat{\includegraphics[width=0.7\textwidth]{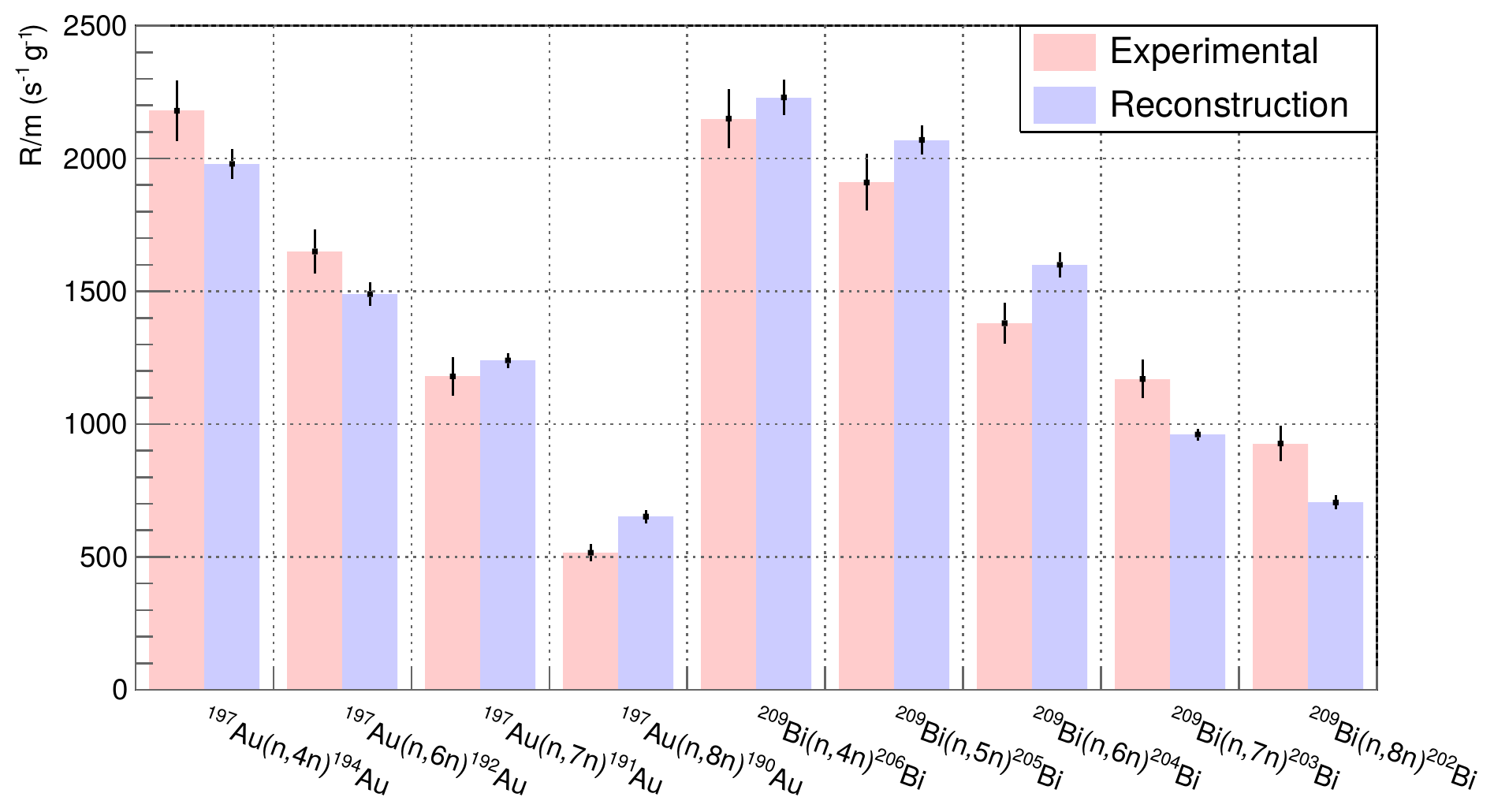}}
\end{center}
\caption{Activation rates per unit mass for all measured reactions. Experimental values are compared to the ones reconstructed after the \textit{minimal} unfolding. }
\label{Fig:ActRate}
\end{figure}

In this analysis, effective cross sections are evaluated from the following libraries: \textsc{Endf/b-vii.1}~\cite{ENDF} for neutron capture (n,$\gamma$) reactions and \textsc{Tendl-2015}~\cite{TENDL} for the threshold reactions. 
This choice is due to the fact that the \textsc{Tendl-2015} library, unlike the \textsc{Endf/b-vii.1}, provides cross section evaluations up to 200~MeV for all the threshold reactions we measured in this experiment. 
In the absence of cross section data above 200~MeV, we perform a linear extrapolation up to 700~MeV. 
Since in the energy region above 100~MeV the cross sections of observed reactions are relatively low and the neutron flux has a decreasing spectrum, we can conclude that this assumption does not significantly affect the final results.

The uncertainties on neutron capture cross section are taken from the BNL-98403-2012-JA Report~\cite{PRITYCHENKO20123120}, that includes a complete list of uncertainties for both thermal cross sections and resonance integrals, based on the Low Fidelity Covariance Project~\cite{LITTLE20082828}.
For threshold reactions, we use the uncertainty evaluations provided by the \textsc{Tendl-2015} library itself. When the uncertainty estimate is not available for some reaction (this happens especially for the (n,xn) reactions involving the emission of many neutrons), we assume 30\% uncertainty, which corresponds to the relative uncertainty quoted for similar reactions.

To study the impact of the cross section uncertainties on the results, we perform a new unfolding, that we will call \textit{minimal}, in which we fix the cross sections to their central values. 
In this way, we get an estimate of the precision that we could reach with this unfolding technique if the cross section data were known with negligible uncertainties.
The results of the \textit{minimal} unfolding are reported in the last column of Tab.~\ref{tab:Results} and prove that most part of the statistical uncertainty evaluated in the \textit{reference} unfolding for the flux groups above 1~MeV is due to the propagation of the cross section uncertainties. 

The \textit{minimal unfolding} is also useful to perform an important \textit{a posteriori} check on the reconstruction of the activation rate data. In this case the cross sections are constrained to the central values reported in literature and can be combined with the resulting flux groups to calculate the activation rates of all the reactions used for the unfolding.
The plots in Fig.~\ref{Fig:ActRate} show the comparison of the experimental activation rates with the ones reconstructed after the \textit{minimal} unfolding. Taking into account that in this analysis we are not including the cross section uncertainties, that can reach up to 30\% for the threshold reactions, we can conclude that the reconstructed activation rates are fully compatible with the measured ones. This result demonstrates that there is no tension between the input data of activation rates and the unfolded neutron flux spectrum.

\subsection{Analysis of systematic uncertainties} \label{sec:syst}

Here we present the analysis of the systematic uncertainties related to the \textit{intra-group} spectrum shapes used to calculate the effective cross sections.
For this purpose, we repeat the unfolding analysis using different spectral shapes in each energy bin.
To identify reasonable shapes to be tested, we take as a benchmark the result of the \textit{reference} unfolding presented in Fig.~\ref{Fig:UnfoldedSpectrum} and we vary the guess spectrum compatibly with the general trend observed in the different energy regions by connecting the points of the flux groups.

\begin{figure}[b!]
\begin{center}
\includegraphics[width=0.9\textwidth]{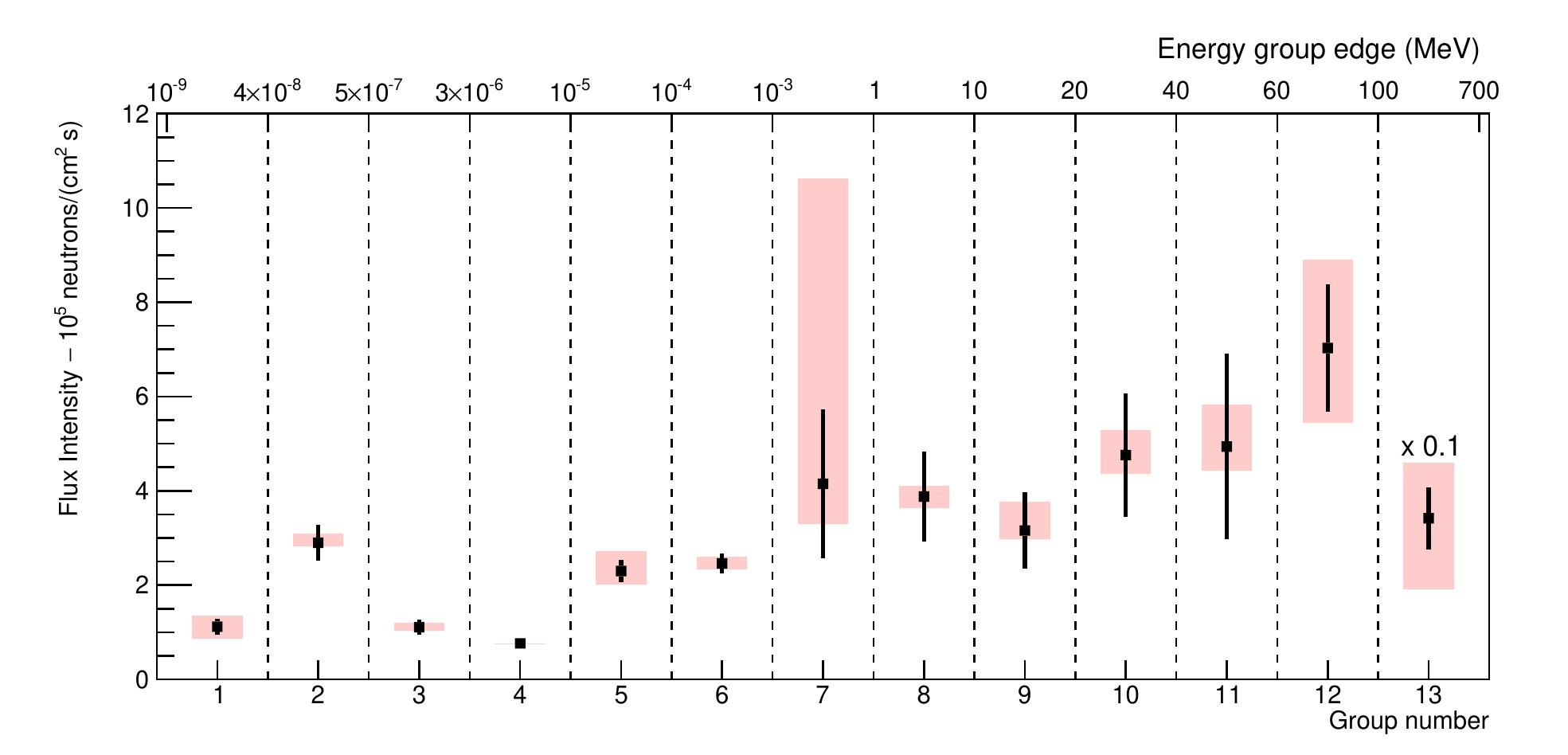}
\end{center}
\caption{Flux intensity for each energy group. Error bars are statistical uncertainties; the colored bands are the ranges obtained in the tests of systematics (using different guess spectra). The intensity of the flux in the last group was divided by 10 for visual purposes.}
\label{Fig:Syst}
\end{figure}

Particularly, we perform the following tests of systematics:
\begin{itemize}
\item For the thermal flux in the first bin, we repeat the unfolding with guess spectra proportional to $E^{-0.5}$ and  $E^{0.5}$, instead of constant;
\item For the flux between $4\times10^{-8}$ and 1~MeV, instead of using a shape proportional to $E^{-1}$, we test guess spectra proportional to $E^{-0.7}$ and  $E^{-1.3}$;
\item For the fast flux between 1 and 100~MeV, we modify the constant flux used in the \textit{reference} unfolding into guess spectra proportional to $E^{-0.5}$ and $E^{0.5}$, respectively;
\item For the last bin, whose intensity was previously constrained to the integral subtended by a spectrum decreasing as $E^{-1}$ and connected with continuity at 100~MeV, we test different shapes and we calculate the corresponding subtended integrals to be used as constraints. In detail, the spectral shapes used for this test are: flux linearly decreasing to zero at $E=700$~MeV; flux $\propto E^{-2}$; flux constant up to $200$~MeV and then $\propto E^{-1}$; flux $\propto E^{-1}$ starting from 50~MeV (instead of 100~MeV used in the \textit{reference} unfolding);
\item Finally, we unfold the neutron spectrum using as a guess the analytic fit of the atmospheric spectrum~\cite{gordon2004measurement} for energies in the range 100~keV--700~MeV; this test is also repeated using the atmospheric guess spectrum only for $E>1$~MeV.
\end{itemize}
In order to get an approximate evaluation of the systematic uncertainty affecting the \textit{reference} unfolding, we analyze the results of all these tests to find the minimum and maximum values of each flux group. 

In Fig.~\ref{Fig:Syst}, we summarize the results of the systematic uncertainty analysis: the black points represent the results of the \textit{reference} unfolding, with the error bars corresponding to the statistical uncertainties; the colored bands are used instead to highlight the range between the minimum and the maximum flux estimates obtained in the tests of systematics.

This analysis shows that the \textit{reference} unfolding results are affected by significant systematic uncertainties only in the energy range from 1~keV to 1~MeV, corresponding to the 7$^{th}$ flux group.
This is not a surprising result since, as previously discussed, the experimental data used to unfold the neutron spectrum do not contain specific signatures to precisely determine the flux in this energy region.
In all the other bins of the spectrum, the systematic uncertainties are not much larger than the statistical ones, and in many bins are even lower.

This analysis proves that the method used to unfold the neutron spectrum allows to determine the results practically independently of the choice of the \textit{intra-group} spectrum shapes, if the experimental data include specific signatures for each bin.\\

Finally, in Tab.~\ref{tab:MacroGroup}, we summarize the results of the neutron flux measurement at ChipIr beamline subdividing the spectrum in 4 macro-groups: 
\begin{itemize}
\item thermal neutrons with $E<0.5$eV;
\item epithermal neutrons in the range [0.5--100]eV;
\item intermediate and fast neutrons in the range [0.1~keV--10~MeV];
\item very fast neutrons with $E>10$MeV;
\end{itemize}
Each macro-group is the sum of two or more groups defined in the \textit{reference} unfolding and its intensity is obtained by summing the corresponding results taking into account the correlations between groups. In this table we also provide a range of systematic uncertainty to be associated to each macro-group. This range is obtained using the results of the unfolding tests presented in this section and, when combined with the statistical uncertainties, allows to get a general picture of the accuracy we could reach for the neutron flux measurement in the different regions of the spectrum.

\begin{table}[h!]
\begin{center}
\begin{tabular}{|lll|c|c|}
\hline
\multicolumn{3}{|c|}{Macro group	} &	Reference unfolding 				&	Range of systematics:	\\
\multicolumn{3}{|c|}{energy range (MeV)} 	&	flux ($10^6$cm$^{-2}$s$^{-1}$)		&	min -- max flux ($10^6$cm$^{-2}$s$^{-1}$)		\\
\hline
$10^{-9}$ & -- & $5\times10^{-7}$	&	0.40 $\pm$ 0.03	&	0.38 -- 0.43	\\
$5\times10^{-7}$ & -- & $10^{-4}$	&	0.42 $\pm$ 0.03	&	0.40 -- 0.45	\\
$10^{-4}$ & -- & 10					&	1.05 $\pm$ 0.17	&	0.95 -- 1.67	\\
10 & -- & 700						&	5.4 $\pm$ 0.7	&	4.2 -- 6.3	\\
\hline
& & Total							&	7.3 $\pm$ 0.7	&	5.9 -- 8.9	\\
\hline
\end{tabular}
\end{center}
\caption{Results of neutron flux measurement at ChipIr with the spectrum subdivided in 4 macro-groups. The results from the \textit{reference} unfolding are shown in column 2. The range of systematic uncertainty reported in the last column is obtained by taking the minimum and the maximum value of each macro-group from the tests with different guess spectra.}
\label{tab:MacroGroup}
\end{table}

%% file: conclusion.tex
The neutron flux of ChipIr, a new fast neutron beamline at the ISIS spallation source, has been measured using multi-foil activation analysis. The method, described in this paper, uses a large selection of activation reactions to cover the wide energy range and to allow the unfolding of the neutron spectrum. \\
A particular attention has been paid to determine the uncertainties, and it is shown that, for the fast neutron part, the main role is played by uncertainties on the tabulated cross sections. \\
The method and results are of general interest for the characterization of beamlines at advanced neutron sources.\\

%% file: main.bbl
\begin{thebibliography}{10}
\expandafter\ifx\csname url\endcsname\relax
  \def\url#1{\texttt{#1}}\fi
\expandafter\ifx\csname urlprefix\endcsname\relax\def\urlprefix{URL }\fi
\expandafter\ifx\csname href\endcsname\relax
  \def\href#1#2{#2} \def\path#1{#1}\fi

\bibitem{russell1995introduction}
G.~Russell, E.~Pitcher, L.~Daemen, Introduction to spallation physics and
  spallation-target design, in: AIP Conference Proceedings, Vol. 346, AIP,
  1995, pp. 93--104.

\bibitem{weisskopf1937statistics}
V.~Weisskopf, Statistics and nuclear reactions, Physical Review 52~(4) (1937)
  295.

\bibitem{wilson1995guided}
C.~Wilson, A guided tour of {ISIS--the UK} spallation neutron source, Neutron
  News 6~(2) (1995) 27--34.

\bibitem{mason2006spallation}
T.~Mason, D.~Abernathy, I.~Anderson, J.~Ankner, T.~Egami, G.~Ehlers,
  A.~Ekkebus, G.~Granroth, M.~Hagen, K.~Herwig, et~al., The spallation neutron
  source in {Oak Ridge}: A powerful tool for materials research, Physica B:
  Condensed Matter 385 (2006) 955--960.

\bibitem{klinkby2016neutron}
E.~B. Klinkby, L.~Zanini, K.~Batkov, F.~Mezei, T.~Schoenfeldt, A.~Takibaev,
  Neutron moderators for the {European Spallation Source}, in: Physics of
  fundamental Symmetries and Interactions (PSI2016), 2016.

\bibitem{lisowski1990alamos}
P.~Lisowski, C.~Bowman, G.~Russell, S.~Wender, The {Los Alamos} national
  laboratory spallation neutron sources, Nuclear Science and Engineering
  106~(2) (1990) 208--218.

\bibitem{guerrero2013performance}
C.~Guerrero, A.~Tsinganis, M.~B. G{\'o}mez~Hornillos,
  {\'A}.~Hern{\'a}ndez~Prieto, A.~Riego~P{\'e}rez, {Performance of the neutron
  time-of-flight facility n TOF at CERN}, European physical journal A 49~(27)
  (2013) 1--15.

\bibitem{prokofiev2009anita}
A.~V. Prokofiev, J.~Blomgren, S.~P. Platt, R.~Nolte, S.~Rottger, A.~N. Smirnov,
  {ANITA}--a new neutron facility for accelerated see testing at the svedberg
  laboratory, in: Reliability Physics Symposium, 2009 IEEE International, IEEE,
  2009, pp. 929--935.

\bibitem{frost2009new}
C.~D. Frost, S.~Ansell, G.~Gorini, {A new dedicated neutron facility for
  accelerated SEE testing at the ISIS facility}, in: Reliability Physics
  Symposium, 2009 IEEE International, IEEE, 2009, pp. 952--955.

\bibitem{blackmore2009development}
E.~W. Blackmore, Development of a large area neutron beam for system testing at
  {TRIUMF}, in: Radiation Effects Data Workshop, 2009 IEEE, IEEE, 2009, pp.
  157--160.

\bibitem{blackmore2014intensity}
E.~W. Blackmore, M.~Trinczek, Intensity upgrade to the {TRIUMF 500 MeV}
  large-area neutron beam, in: Radiation Effects Data Workshop (REDW), 2014
  IEEE, IEEE, 2014, pp. 1--5.

\bibitem{wender2016neutron}
S.~A. Wender, Neutron-induced failures in semiconductor devices, Tech. rep.,
  Los Alamos National Lab.(LANL), Los Alamos, NM (United States) (2016).

\bibitem{prokofiev2013cup}
A.~V. Prokofiev, E.~Passoth, A.~Hjalmarsson, M.~Majerle, {CUP—A new high-flux
  irradiation position at the ANITA neutron facility at TSL}, in: Radiation and
  Its Effects on Components and Systems (RADECS), 2013 14th European Conference
  on, IEEE, 2013, pp. 1--8.

\bibitem{andreani2008facility}
C.~Andreani, A.~Pietropaolo, A.~Salsano, G.~Gorini, M.~Tardocchi,
  A.~Paccagnella, S.~Gerardin, C.~Frost, S.~Ansell, S.~Platt, Facility for fast
  neutron irradiation tests of electronics at the isis spallation neutron
  source, Applied Physics Letters 92~(11) (2008) 114101.

\bibitem{bedogni2009characterization}
R.~Bedogni, A.~Esposito, C.~Andreani, R.~Senesi, M.~P. De~Pascale, P.~Picozza,
  A.~Pietropaolo, G.~Gorini, C.~D. Frost, S.~Ansell, Characterization of the
  neutron field at the {ISIS-VESUVIO} facility by means of a bonner sphere
  spectrometer, Nuclear Instruments and Methods in Physics Research Section A:
  Accelerators, Spectrometers, Detectors and Associated Equipment 612~(1)
  (2009) 143--148.

\bibitem{cazzaniga2015telescope}
C.~Cazzaniga, M.~Rebai, M.~Tardocchi, G.~Croci, M.~Nocente, S.~Ansell,
  C.~Frost, G.~Gorini, A telescope proton recoil spectrometer for fast neutron
  beam-lines, Progress of Theoretical and Experimental Physics 2015~(7) (2015)
  073H01.

\bibitem{wender1993fission}
S.~Wender, S.~Balestrini, A.~Brown, R.~Haight, C.~Laymon, T.~Lee, P.~Lisowski,
  W.~McCorkle, R.~Nelson, W.~Parker, et~al., A fission ionization detector for
  neutron flux measurements at a spallation source, Nuclear Instruments and
  Methods in Physics Research Section A: Accelerators, Spectrometers, Detectors
  and Associated Equipment 336~(1-2) (1993) 226--231.

\bibitem{cazzaniga2016characterization}
C.~Cazzaniga, C.~Frost, T.~Minniti, E.~Schooneveld, E.~P. Cippo, M.~Tardocchi,
  M.~Rebai, G.~Gorini, Characterization of the high-energy neutron beam of the
  {PRISMA} beamline using a diamond detector, Journal of Instrumentation
  11~(07) (2016) P07012.

\bibitem{rebai2016time}
M.~Rebai, A.~Fazzi, C.~Cazzaniga, G.~Croci, M.~Tardocchi, E.~P. Cippo,
  C.~Frost, D.~Zaccagnino, V.~Varoli, G.~Gorini, Time-stability of a
  single-crystal diamond detector for fast neutron beam diagnostic under alpha
  and neutron irradiation, Diamond and Related Materials 61 (2016) 1--6.

\bibitem{cazzaniga2017charge}
C.~Cazzaniga, M.~Rebai, J.~G. Lopez, M.~Jimenez-Ramos, M.~Girolami, D.~Trucchi,
  A.~Bellucci, C.~Frost, M.~Garcia-Munoz, M.~Nocente, et~al., Charge collection
  uniformity and irradiation effects of synthetic diamond detectors studied
  with a proton micro-beam, Nuclear Instruments and Methods in Physics Research
  Section B: Beam Interactions with Materials and Atoms 405 (2017) 1--10.

\bibitem{croci2015gem}
G.~Croci, G.~Claps, C.~Cazzaniga, L.~Foggetta, A.~Muraro, P.~Valente,
  {GEM-based} detectors for thermal and fast neutrons, The European Physical
  Journal Plus 130~(6) (2015) 118.

\bibitem{cazzaniga2014thin}
C.~Cazzaniga, M.~Nocente, M.~Tardocchi, A.~Fazzi, A.~Hjalmarsson, D.~Rigamonti,
  G.~Ericsson, G.~Gorini, {Thin YAP: Ce and LaBr 3: Ce scintillators as proton
  detectors of a thin-film proton recoil neutron spectrometer for fusion and
  spallation sources applications}, Nuclear Instruments and Methods in Physics
  Research Section A: Accelerators, Spectrometers, Detectors and Associated
  Equipment 751 (2014) 19--22.

\bibitem{AbsoluteFlux}
A.~Borio~di Tigliole, A.~Cammi, D.~Chiesa, M.~Clemenza, S.~Manera, M.~Nastasi,
  L.~Pattavina, R.~Ponciroli, S.~Pozzi, M.~Prata, E.~Previtali, A.~Salvini,
  M.~Sisti, {TRIGA} reactor absolute neutron flux measurement using activated
  isotopes, Progress in Nuclear Energy 70 (2014) 249--255.

\bibitem{FluxDistribution}
D.~Chiesa, M.~Clemenza, M.~Nastasi, S.~Pozzi, E.~Previtali, G.~Scionti,
  M.~Sisti, M.~Prata, A.~Salvini, A.~Cammi, Measurement and simulation of the
  neutron flux distribution in the {TRIGA Mark~II} reactor core, Annals of
  Nuclear Energy 85 (2015) 925 -- 936.

\bibitem{BayesianSpectrum}
D.~Chiesa, E.~Previtali, M.~Sisti, Bayesian statistics applied to neutron
  activation data for reactor flux spectrum analysis, Annals of Nuclear Energy
  70 (2014) 157 -- 168.

\bibitem{JAGS}
M.~Plummer, JAGS Version 3.3.0 User Manual (2012).

\bibitem{Gelman}
A.~Gelman, J.~B. Carlin, H.~S. Stern, D.~B. Rubin, Bayesian Data Analysis,
  Chapman and Hall/CRC, 2004.

\bibitem{Shieldwerx}
Shieldwerx, {LLC}, \url{http://www.shieldwerx.com}, accessed: 2018-03-01.

\bibitem{GoodFellow}
Goodfellow {Cambridge Ltd.}, \url{http://www.goodfellow.com}, accessed:
  2018-03-01.

\bibitem{FestaIMAT}
G.~Festa, C.~Andreani, L.~Arcidiacono, G.~Burca, W.~Kockelmann, T.~Minniti,
  R.~Senesi, Characterization of $\gamma$-ray background at {IMAT} beamline of
  {ISIS Spallation Neutron Source}, Journal of Instrumentation 12~(08) (2017)
  P08005.

\bibitem{PIETROPAOLO2006826}
A.~Pietropaolo, M.~Tardocchi, E.~Schooneveld, R.~Senesi, Characterization of
  the $\gamma$ background in epithermal neutron scattering measurements at
  pulsed neutron sources, Nuclear Instruments and Methods in Physics Research
  Section A: Accelerators, Spectrometers, Detectors and Associated Equipment
  568~(2) (2006) 826 -- 838.

\bibitem{C4JA00183D}
A.~Miceli, G.~Festa, R.~Senesi, E.~P. Cippo, L.~Giacomelli, M.~Tardocchi,
  A.~Scherillo, E.~Schooneveld, C.~Frost, G.~Gorini, C.~Andreani, Measurements
  of gamma-ray background spectra at spallation neutron source beamlines, J.
  Anal. At. Spectrom. 29 (2014) 1897--1903.

\bibitem{GEANT4}
S.~Agostinelli, {\it et al.}, $\textsc{Geant4}$: A simulation toolkit, Nucl.
  Instr. Meth. A 506 (2003) 250--303.

\bibitem{PDG2016}
C.~Patrignani, {Particle~Data~Group}, {Review of Particle Physics}, Chinese
  Physics C 40~(10) (2016) 100001.

\bibitem{PlattSpectra}
S.~P. Platt, A.~V. Prokofiev, X.~X. Cai, Fidelity of energy spectra at neutron
  facilities for single-event effects testing, in: 2010 IEEE International
  Reliability Physics Symposium, 2010, pp. 411--416.

\bibitem{mcnp}
{X-5 Monte Carlo Team}, {MCNP - A General Monte Carlo N-Particle Transport
  Code, Version~5}, Los Alamos National Laboratory (2008).

\bibitem{ENDF}
M.~B. Chadwick, {\it et al.}, {ENDF/B-VII.1 Nuclear Data for Science and
  Technology: Cross Sections, Covariances, Fission Product Yields and Decay
  Data}, Nuclear Data Sheets 112 (2011) 2887--2996.

\bibitem{TENDL}
A.~Koning, D.~Rochman, {Modern Nuclear Data Evaluation with the TALYS Code
  System}, Nuclear Data Sheets 113~(12) (2012) 2841 -- 2934, special Issue on
  Nuclear Reaction Data.

\bibitem{PRITYCHENKO20123120}
B.~Pritychenko, S.~Mughabghab, {Neutron Thermal Cross Sections, Westcott
  Factors, Resonance Integrals, Maxwellian Averaged Cross Sections and
  Astrophysical Reaction Rates Calculated from the ENDF/B-VII.1, JEFF-3.1.2,
  JENDL-4.0, ROSFOND-2010, CENDL-3.1 and EAF-2010 Evaluated Data Libraries},
  Nuclear Data Sheets 113~(12) (2012) 3120 -- 3144, special Issue on Nuclear
  Reaction Data.

\bibitem{LITTLE20082828}
R.~Little, T.~Kawano, G.~Hale, M.~Pigni, M.~Herman, P.~Oblo\v{z}insk\'{y},
  M.~Williams, M.~Dunn, G.~Arbanas, D.~Wiarda, R.~McKnight, J.~McKamy,
  J.~Felty, {Low-fidelity Covariance Project}, Nuclear Data Sheets 109~(12)
  (2008) 2828 -- 2833, special Issue on Workshop on Neutron Cross Section
  Covariances June 24-28, 2008, Port Jefferson, New York, USA.

\bibitem{gordon2004measurement}
M.~Gordon, P.~Goldhagen, K.~Rodbell, T.~Zabel, H.~Tang, J.~Clem, P.~Bailey,
  Measurement of the flux and energy spectrum of cosmic-ray induced neutrons on
  the ground, IEEE Transactions on Nuclear Science 51~(6) (2004) 3427--3434.

\end{thebibliography}
